\documentclass[11pt]{article}

\usepackage{latexsym,amsmath,amssymb}
\usepackage{epsfig}
\usepackage{tabls}
\usepackage{times,mathptmx}
\usepackage{tabularx}
\usepackage{xspace}
\usepackage{algorithm, algorithmic}
\usepackage{graphicx}

\usepackage{typearea}
\typearea{15}

\newcommand{\ignore}[1]{}

\ignore{

\makeatletter
 \addtolength{\textheight}{20pt}
 \addtolength{\footskip}{-20pt}
\makeatother

}

\newtheorem{thm}{Theorem}
\newtheorem{theorem}[thm]{Theorem}

\newtheorem{lemma}[thm]{Lemma}
\newtheorem{corollary}[thm]{Corollary}
\newtheorem{claim}[thm]{Claim}
\newtheorem{definition}{Definition}
\newtheorem{example}{Example}

\def\sse{\subseteq}

\def\opt{{\sf Opt}\xspace}
\def\lat{{\sf Lat}\xspace}
\def\tsp{{\sf Tsp}\xspace}

\def\is{\mathcal{I}}
\def\js{\ensuremath{\mathcal{J}}}

\def\ds{\mathcal{D}}

\def\iso{\ensuremath{{\sf IsoAlg}}\xspace}
\def\palg{\ensuremath{{\sf Partition}}\xspace}
\def\palgl{\ensuremath{{\sf PartnLat}}\xspace}

\def\odt{{\sf ODT}\xspace}

\newcommand{\profit}{\phi}

\def\latency{{\sf latency}}

\def\E{\mathbf{E}}

\def\isoprob{\ensuremath{{\sf Isolation}}\xspace}
\def\isotime{\ensuremath{{\sf IsoTime}}}
\def\isoopt{{\isotime}^*}

\def\stsp{\ensuremath{{\sf AdapTSP}}\xspace}
\def\strp{\ensuremath{{\sf AdapTRP}}\xspace}

\def\lpgst{\ensuremath{{\sf LPGST}}\xspace}

\def\gso{\ensuremath{{\sf GSO}}\xspace}
\def\lgs{{\sf LGST}\xspace}
\def\mgs{{\sf GST}\xspace}

\def\dtp{optimal decision tree problem\xspace}

\def\ts{\textstyle}

\newcommand{\initOneLiners}{%
    \setlength{\itemsep}{0pt}
    \setlength{\parsep }{0pt}
    \setlength{\topsep }{0pt}
}

\newenvironment{pf}{

\noindent{\bf Proof:}} {\hfill$\blacksquare$

}

\usepackage{url}
\makeatletter
\def\url@leostyle{%
  \@ifundefined{selectfont}{\def\UrlFont{\sf}}{\def\UrlFont{\small\ttfamily}}}
\makeatother
\title{Approximation Algorithms for Optimal Decision Trees  \\ and Adaptive
  TSP Problems}
\author{
Anupam Gupta\thanks{Computer Science Department, Carnegie Mellon
    University.}  \and Viswanath Nagarajan\thanks{ Industrial and Operations Engineering Department, University of Michigan.} \and  R. Ravi\thanks{Tepper School of Business, Carnegie Mellon University.}}
\date{}

\begin{document}

\maketitle
\begin{abstract}
We consider the problem of constructing {\em optimal decision trees}: 
  given a collection of tests which can disambiguate between a set of
  $m$ possible diseases, each test having a cost, and the a-priori
  likelihood of any particular disease, what is a
  good adaptive strategy to perform these tests to minimize the expected
  cost to identify the disease?
This problem has been studied in several works, with $O(\log m)$-approximations known in the special cases when either
costs or probabilities are uniform. In this paper, we settle the approximability of the general problem by giving a tight $O(\log m)$-approximation
algorithm.

We also consider a substantial generalization, the
{\em adaptive traveling salesman problem}. Given an underlying metric space, a random subset $S$ of vertices is drawn
from a known distribution, but $S$ is initially unknown---we get information about whether any vertex is in $S$
only when it is visited. What is a good adaptive strategy to visit all vertices in the random subset $S$ while
minimizing the expected distance traveled? This problem has applications in routing message ferries in ad-hoc networks,
and also models switching costs between tests in the optimal decision tree problem. We give a poly-logarithmic
approximation algorithm for adaptive TSP, which is nearly best possible due to a connection to the well-known group Steiner tree problem. Finally, we consider the related {\em adaptive traveling repairman
problem}, where the goal is to compute an adaptive tour minimizing the expected sum of arrival times of vertices in the random subset $S$; we obtain a poly-logarithmic approximation algorithm for this problem as well.
\end{abstract}


\section{Introduction}

Consider the following two adaptive covering optimization problems:
\begin{itemize}

\item \emph{Adaptive TSP under stochastic demands} (\stsp). A traveling  salesperson is given a metric space $(V,d)$ and distinct subsets $S_1,
  S_2, \ldots, S_m \sse V$ such that $S_i$ appears with probability  $p_i$ (and $\sum_i p_i = 1$). She needs to serve requests at a random
subset $S$ of locations drawn from this distribution. However, she does not know the identity of the
  random subset: she can only visit locations, at which time she finds
  out whether or not that location is part of the subset $S$. What adaptive
  strategy should she use to minimize the expected time to serve all
  requests in the random  set $S$?

\item \emph{Optimal Decision Trees.} Given a set of $m$ diseases, there
  are $n$ binary tests that can be used to disambiguate between these
  diseases.  If the cost of performing test $t \in [n]$ is $c_t$,
  and we are given the likelihoods $\{p_j\}_{j \in [m]}$ that a typical
  patient has the disease $j$, what (adaptive) strategy should the
  doctor use for the tests to minimize the expected cost to identify the
  disease?   
  \end{itemize}

It can be shown that the \dtp is a special case of the adaptive TSP problem: a formal reduction is given in
Section~\ref{sec:odt}.  In both these problems we want to devise adaptive strategies, which take into account
the revealed information in the queries so far (e.g., locations already visited, or tests already done) to determine
the future course of action. Such an {\em adaptive solution} corresponds naturally to a decision tree, where nodes encode the current ``state'' of the solution and branches represent observed random outcomes: see Definition~\ref{def:dt} for a formal definition.  A simpler class of solutions, that have been useful in some other adaptive optimization problems, eg.~\cite{DGV08,GuhaM09,BGLMNR12}, are {\em non-adaptive solutions}, which are specified by just an ordered list of actions. However there are instances for both the above problems where the optimal adaptive solution costs much less than the optimal non-adaptive solution.  
Hence it is
essential that we find good  adaptive solutions.

The \emph{\dtp} has long been studied, its NP-hardness was shown by Hyafil and Rivest in 1976~\cite{rivest-hyafil}  and
many references and applications can be found in~\cite{N11}. 
There have been a large number of papers providing algorithms for this problem~\cite{garey-graham,loveland,kmb,dasgupta,AH12,CPRAM11,N11,gb09}. The best results yield approximation ratios of $O\left(\log \frac1{p_{\min}}\right)$ and $O\left(\log (m\,\frac{c_{max}}{c_{min}})\right)$, where $p_{min}$ is the minimum non-zero probability and $c_{max}$ (resp. $c_{min}$) is the maximum (resp. minimum) cost. In the special cases when the likelihoods $\{p_j\}$ or the costs
$\{c_t\}$ are all polynomially bounded in $m$, these imply an $O(\log m)$-approximation
algorithm. However, there are instances (when probabilities and costs are exponential) demonstrating an $\Omega(m)$ approximation guarantee for all previous algorithms. On the hardness side, an $\Omega(\log m)$ hardness of approximation (assuming $P\ne NP$) is known for the \dtp~\cite{CPRAM11}. 
While the existence of an $O(\log m)$-approximation algorithm for the general \dtp has been posed as an open question,
it has not been answered prior to this work.

Optimal decision tree is also a basic problem in {\em average-case active learning}~\cite{dasgupta,N11,gb09}. In this
application, there is a set of $n$ data points, each of which is associated with a $+$ or $-$ label. The labels are
initially unknown. A classifier is a partition of the data points into $+$ and $-$ labels. The true classifier $h^*$ is
the partition corresponding to the actual data labels. The learner knows beforehand, a ``hypothesis class'' $H$
consisting of $m$ classifiers; it is assumed that the true classifier $h^*\in H$. Furthermore, in the average case
model, there is a known distribution $\pi$ of $h^*$ over $H$. The learner wants to determine $h^*$ by querying labels
at various points. There is a cost $c_t$ associated with querying the label of each data point $t$. An active learning
strategy involves adaptively querying labels of data points until $h^*\in H$ is identified. The goal is to compute a
strategy that minimizes the expectation (over $\pi$) of the cost of all queried points. This is precisely the \dtp,
with points being tests and classifiers corresponding to diseases.

\medskip
Apart from being a natural adaptive routing problem, \stsp has many applications in the setting of message ferrying in
ad-hoc networks~\cite{ZA03,SRJB03,ZAZ04,ZAZ05,HLS10}. We cite two examples below:
\begin{itemize}
\item {\em Data collection in sparse sensor networks (see eg.~\cite{SRJB03}).} A collection of sensors is spread over a large geographic area, and
one needs to periodically gather sensor data at a base station. Due to the power and cost overheads of setting up a
communication network between the sensors, the data collection is instead performed by a mobile device (the message
ferry) that travels in this space from/to the base station. On any given day, there is a known distribution ${\cal D}$
of the subset $S$ of sensors that contain new information: this might be derived from historical data or domain
experts. The routing problem for the ferry then involves computing a tour (originating from the base station) that
visits all sensors in $S$, at the minimum expected cost.

\item {\em Disaster management (see eg.~\cite{ZAZ04}).} Consider a post-disaster situation, in which usual
communication networks have broken down. In this case, vehicles can be used in order
to visit locations and assess the damage. Given a distribution of the set of affected locations, the goal here is to
route a vehicle that visits all affected locations as quickly as possible  in expectation.
\end{itemize}
In both these applications, due to the absence of a direct communication network, the information at any location is
obtained only when it is visited: this is precisely the \stsp problem.

\subsection{Our Results and Techniques}
In this paper, we settle the approximability of the \dtp:
\begin{theorem}
  \label{thm:main1}
  There is an $O(\log m)$-approximation algorithm for the \dtp with arbitrary test
  costs and arbitrary probabilities, where $m$ is the number of diseases. The problem admits the same
  approximation ratio even when the tests have non-binary outcomes.
\end{theorem}
In fact, this result arises as a special case of the following theorem:
\begin{theorem}
  \label{thm:main2}
  There is an $O(\log^2 n \log m)$-approximation algorithm for
  the adaptive Traveling Salesman Problem, where $n$ is the number of vertices and $m$ the number of scenarios in the
  demand distribution.
\end{theorem}

To solve the \stsp problem, we first solve the ``isolation problem'', which seeks to identify which of the $m$
scenarios has materialized. Once we know the scenario we can visit its vertices using any constant-factor approximation algorithm for TSP. 
The high-level idea behind our algorithm for the isolation problem is this---suppose each vertex lies in at most half the scenarios; then if we
visit one vertex in each of the $m$ scenarios using a short tour, which is an instance of the {\em group Steiner tree} problem\footnote{In the group Steiner tree problem~\cite{gkr} the input is a metric $(V,d)$ with root $r\in V$ and groups $\{X_i\sse V\}$ of vertices; the goal is to compute a minimum length tour originating from $r$ that visits at least one vertex of each group.}, we'd
notice at least one of these vertices to have a demand; this would reduce the number of possible scenarios by at least $50\%$ and we can recursively run the algorithm on the remaining scenarios.
This is an over-simplified view, and there are many details to handle: we need not visit all
scenarios---visiting all but one allows us to infer the last one by exclusion; the expectation in the objective
function means we need to solve a \emph{minimum-sum} version of group Steiner tree; not all vertices need lie in less
than half the scenarios.  Another major issue is that we do not want our performance to depend on the magnitude of the
probabilities, as some of them may be exponentially small.  
Finally, we need to charge our cost directly against the optimal decision tree. 
All these issues can indeed be resolved to obtain Theorem~\ref{thm:main2}.

The algorithm for the isolation problem involves an interesting combination of ideas from the group Steiner~\cite{gkr,ccgg} and minimum
latency TSP~\cite{bccprs,cgrt,fhr} problems---it uses a greedy approach that is greedy with respect to two different
criteria, namely the probability measure and the number of scenarios. This idea is formalized in our algorithm for 
the {\em partial latency} group Steiner (\lpgst) problem, which is a key subroutine for \isoprob. While this \lpgst
problem is harder to approximate than the standard group Steiner tree (see Section~\ref{sec:prelim}), for which $O(\log^2n \log m)$ is the best approximation ratio, we show  that it admits a better
$\left(O(\log^2n),\,4\right)$ {\em bicriteria} approximation algorithm. Moreover, even this bicriteria approximation
guarantee for \lpgst suffices to obtain an $O(\log^2n\cdot \log m)$-approximation algorithm for \isoprob.

We also show that both \stsp and the isolation problem are $\Omega(\log^{2-\epsilon} n)$ hard to approximate even on tree metrics;
our results are essentially best possible on such metrics, and we lose an extra logarithmic factor to go to general
metrics, as in the group Steiner tree problem. Moreover, any improvement to the result in Theorem~\ref{thm:main2} would lead to a similar improvement for the group Steiner tree problem~\cite{gkr,hk03,cp05} which is a long-standing open question. 

For the \dtp, we show that we can use a variant of minimum-sum set cover~\cite{FLT04} which is the special case of \lpgst on star-metrics. This avoids an $O(\log^2 n)$ loss in the approximation guarantee, and hence gives us an $O(\log
m)$-approximation algorithm which is best possible~\cite{CPRAM11}. Although this  variant of min-sum set cover is $\Omega(\log m)$-hard to approximate (it
generalizes set cover as shown in Section~\ref{sec:prelim}), we again give a constant factor bicriteria approximation algorithm, which leads to the $O(\log
m)$-approximation for optimal decision tree. Our result further reinforces the close connection between the min-sum set
cover problem and the \dtp that was first noticed by~\cite{CPRAM11}.

Finally, we consider the related adaptive traveling repairman problem (\strp), which has the same input as \stsp, but
the objective is to minimize the expected sum of arrival times at vertices in the materialized demand set. In this setting, we
cannot first isolate the scenario and then visit all its nodes, since a long isolation tour may negatively impact the
arrival times. So \strp (unlike \stsp) cannot be reduced to the isolation problem. However, we show that our techniques for \stsp
are robust, and can be used to obtain:
\begin{theorem}
  \label{thm:main3}
  There is an $O(\log^2 n \log m)$-approximation algorithm for
  the adaptive Traveling Repairman Problem, where $n$ is the number of vertices and $m$ the number of scenarios in the
  demand distribution.
\end{theorem}

\medskip
\noindent {\bf Paper Outline:} The results on the isolation problem appear in Section~\ref{sec:iso}. We obtain the improved approximation algorithm for
optimal decision tree in Section~\ref{sec:odt}. The algorithm for the adaptive traveling salesman problem is in Section~\ref{sec:stsp}; Appendix~\ref{app:hardness} contains a nearly matching  hardness of approximation result. Finally, Section~\ref{sec:strp} is on the adaptive traveling repairman problem.


\subsection{Other Related Work}
The \dtp has been studied earlier by many authors, with algorithms and hardness results being shown
by~\cite{garey-graham,rivest-hyafil,loveland,kmb,AH12,dasgupta,CPRAM11,cprs09,gb09}. As mentioned above, the algorithms
in these papers gave $O(\log m)$-approximation ratios only when the probabilities or costs (or both) are
polynomially-bounded. The early papers on optimal decision tree considered tests with only binary outcomes.
More recently,~\cite{CPRAM11} studied the generalization with $K\ge 2$ outcomes per test, and gave an $O(\log
K\cdot \log m)$-approximation under uniform costs. Subsequently,~\cite{cprs09} improved this bound to $O(\log m)$,
again under uniform costs. Later, \cite{gb09} gave an algorithm under arbitrary costs and
probabilities, achieving an approximation ratio of $O\left(\log \frac1{p_{\min}}\right)$ or $O\left(\log
(m\,\frac{c_{max}}{c_{min}})\right)$. This is the previous best approximation guarantee; see also Table~1
in~\cite{gb09} for a summary of these results. We note that in terms of the number $m$ of diseases, the previous best approximation guarantee is only $\Omega(m)$. On the other hand, there is an $\Omega(\log m)$ hardness of approximation for the
\dtp~\cite{CPRAM11}. Our $O(\log m)$-approximation algorithm for arbitrary costs and probabilities solves an open problem from
these papers. A crucial aspect of this algorithm is that it is   non greedy. All previous results were based on
variants of a greedy algorithm.

There are many results on adaptive optimization dealing with covering problems. E.g., \cite{goemansv06} considered the adaptive set-cover problem; they
gave an $O(\log n)$-approximation when sets may be chosen multiple times, and an $O(n)$-approximation when each set may
be chosen at most once. The latter approximation ratio was improved in~\cite{msw07} to $O(\log^2 n\,\log m)$, and subsequently to the
best-possible $O(\log n)$-approximation ratio by~\cite{lpry08}, also using a greedy algorithm.  In recent
work~\cite{GK11} generalized adaptive set-cover to a setting termed `adaptive submodularity', and gave many
applications.  In all these problems, the adaptivity-gap  (ratio between optimal adaptive and non-adaptive solutions) is large, as is the case for the problems considered in this paper, and so the solutions
need to be inherently adaptive.

The \stsp problem is related to universal TSP~\cite{jlnrs,ghr} and {\em
  a priori} TSP~\cite{jaillet,ss08,st08} only in spirit---in both the
universal and \emph{a priori} TSP problems, we seek a master tour which is shortcut once the demand set is known, and
the goal is to minimize the worst-case or expected length of the shortcut tour. The crucial difference is that the
demand subset is revealed {\em in toto} in these two problems, leaving no possibility of adaptivity---this is in
contrast to the slow revelation of the demand subset that occurs in \stsp.


\section{Preliminaries}\label{sec:prelim}
We work with a  finite metric $(V,d)$ that is given by a set $V$ of $n$ vertices and distance function $d:V\times V\rightarrow \mathbb{R}_+$. 
As usual, we assume that the distance function is symmetric and satisfies the triangle inequality. For any
integer $t\ge 1$, we let $[t]:=\{1,2,\ldots,t\}$.

\begin{definition}[$r$-tour]
Given a metric $(V,d)$ and vertex $r\in V$, an \emph{$r$-tour} is any sequence $(r=u_0,u_1,\cdots,u_k=r)$ of vertices that  begins and ends at $r$. The length of such an $r$-tour is $\sum_{i=1}^k d(u_{i},u_{i-1})$, the total length of all edges in the tour. 
\end{definition}

Throughout this paper, we deal with demand distributions over vertex-subsets that are specified explicitly. A demand
distribution $\ds$ is specified by $m$ distinct subsets $\{S_i\sse V\}_{i=1}^m$ having associated probabilities
$\{p_i\}_{i=1}^m$ such that $\sum_{i=1}^m p_i=1$. This means that the realized subset $D\sse V$ of demand-vertices will
always be one of $\{S_i\}_{i=1}^m$, where $D=S_i$ with probability $p_i$ (for all $i\in [m]$). We also refer to the
subsets $\{S_i\}_{i=1}^m$ as {\em scenarios}. The following definition captures adaptive strategies.

\smallskip
\begin{definition}[Decision Tree]
  \label{def:dt}
  A decision tree $T$ in metric $(V,d)$ is a
  rooted binary tree where each non-leaf node of $T$ is labeled with a vertex $u\in V$,
    and its two children $u_{yes}$ and $u_{no}$ correspond to the
    subtrees taken if there \underline{is} demand at $u$ or if there is \underline{no} demand at $u$.
Thus given any realized demand $D\sse V$, a unique path $T_{D}$ is followed in $T$ from the root down to a leaf.
\end{definition}
Depending on the problem under consideration, there are additional constraints on decision tree $T$ and the
\emph{expected cost} of $T$ is also suitably defined.  There is a (problem-specific) cost $C_i$ associated with each
scenario $i\in [m]$ that depends on path $T_{S_i}$, and the expected cost of $T$ (under distribution $\ds$) is then
$\sum_{i=1}^m p_i\cdot C_i$. For example in \stsp, cost $C_i$ corresponds to the length of path $T_{S_i}$.

\smallskip

Since we deal with explicitly specified demand distributions $\ds$, all decision trees we consider will have size polynomial in $m$ (support size of $\ds$) and $n$ (number of vertices).

\paragraph{Adaptive Traveling Salesman}
This problem consists of a metric $(V,d)$ with root $r\in V$ and a demand distribution $\ds$ over subsets of
vertices. The  information on whether or not there is demand at a vertex $v$ is
obtained only when that vertex $v$ is visited. The objective is to find an adaptive strategy that minimizes the expected
time to visit all vertices of the realized scenario drawn from $\ds$.

We assume that the distribution $\ds$ is specified {\em explicitly} with a support-size of $m$.
This allows us to model demand distributions that are arbitrarily correlated across vertices. We note however that the
running time and performance of our algorithm will depend on the support size. The most general setting would be to consider black-box access to the distribution $\ds$: however, as shown
in~\cite{Vish-thesis}, in this setting there is no $o(n)$-approximation algorithm for \stsp  
that uses a polynomial number of samples from the distribution.
 One could also consider \stsp under independent demand distributions. In this case there is a trivial constant-factor approximation algorithm, that visits all vertices having non-zero probability along an approximately minimum TSP tour; note that any feasible solution must visit all vertices with non-zero probability as otherwise (due to the independence assumption) there would be a positive probability of not satisfying a demand. 
 
\begin{definition}[Adaptive TSP]\label{def:stsp} The input is a metric $(V,d)$, root $r\in V$ and demand distribution $\ds$
given by $m$ distinct subsets $\{S_i\sse V\}_{i=1}^m$ with probabilities
$\{p_i\}_{i=1}^m$ (where $\sum_{i=1}^m p_i=1$). The goal in \stsp 
is to compute a decision tree $T$ in metric $(V,d)$ such that:
  \begin{itemize}
  \item the root of $T$ is labeled with the root vertex $r$, and
  \item for each scenario $i\in[m]$, the path $T_{S_i}$ followed on input $S_i$
   contains \underline{all} vertices in $S_{i}$.
  \end{itemize}
  The objective function is to minimize the expected tour length
  $\sum_{i=1}^m p_i \cdot d(T_{S_i})$, where $d(T_{S_i})$ is
  the length of the tour that starts at $r$, visits the vertices on path
  $T_{S_i}$ in that order, and returns to $r$.
\end{definition}

\paragraph{Isolation Problem}
This is closely related to \stsp. The input is  the same as \stsp, but the goal
is just to identify the unique scenario that 
has materialized, and not to visit all the vertices in the realized scenario.
\smallskip
\begin{definition}[Isolation Problem]\label{def:iso}
Given metric $(V,d)$, root $r$ and demand distribution $\ds$, the goal in \isoprob is to compute a decision tree $T$ in
metric $(V,d)$ such that:
  \begin{itemize}
  \item the root of $T$ is labeled with the root vertex $r$, and
  \item for each scenario $i\in[m]$, the path $T_{S_i}$ followed on input $S_i$ ends at a \underline{distinct} leaf-node of $T$.
    \end{itemize}
The objective  is to
  minimize the expected tour length $\isotime(T):=\sum_{i=1}^m p_i \cdot
  d(T_{S_i})$, where $d(T_{S_i})$ is the length of the $r$-tour that visits the vertices on path $T_{S_i}$ in that order, and returns to $r$.
\end{definition}
\smallskip

The only difference between \isoprob and \stsp is that the tree path $T_{S_i}$ in \isoprob need not contain
all vertices of $S_i$, and the paths for different scenarios must  end at distinct leaf-nodes. In Section~\ref{sec:stsp} we show that  any approximation algorithm for \isoprob leads to an approximation algorithm for \stsp. So we focus on designing algorithms for \isoprob.

\paragraph{Optimal Decision Tree} This problem involves identifying a random disease from a set of possible diseases using binary tests.  

\smallskip
\begin{definition}[Optimal Decision Tree]\label{def:odt}
The input is a set of $m$ diseases with 
probabilities $\{p_i\}_{i=1}^m$ that sum to one, and a collection $\{T_j\sse[m]\}_{j=1}^n$ of $n$ binary tests with
 costs $\{c_j\}_{j=1}^n$. There is exactly one realized disease: each disease $i\in[m]$ occurs with probability $p_i$. Each test $j\in [n]$ returns a positive outcome for subset $T_j$ of  diseases and returns a negative outcome for the rest $[m]\setminus T_j$. The goal in \odt is to compute a decision tree $Q$ where each internal node is  labeled by a test and has two children corresponding to  positive/negative test outcomes, such that for each $i\in[m]$ the path $Q_i$ followed under disease $i$ ends at a distinct leaf node of $Q$. The objective is to minimize the expected cost $\sum_{i=1}^m p_i\cdot c(Q_i)$ where $c(Q_i)$ is the sum of test-costs along path $Q_i$.
\end{definition}
\smallskip

Notice that the optimal decision tree problem is exactly \isoprob on a weighted star metric. Indeed, given an instance of \odt, consider a metric $(V,d)$ induced by a weighted star with  center $r$ and $n$ leaves corresponding to the tests. For each $j\in
  [n]$, we set $d(r,j)=\frac{c_j}2$. The demand scenarios are as
  follows: for each $i\in [m]$ scenario $i$ has demands $S_i=\{j\in [n]\mid i\in T_j\}$.
  It is easy to see that this \isoprob instance corresponds
  exactly to the optimal decision tree instance. See  Section~\ref{sec:odt} for an example. So any algorithm for  \isoprob on star-metrics can be used to solve \odt as well.

\paragraph{Useful Deterministic Problems} Recall that the group Steiner tree problem~\cite{gkr,hk03} consists of a metric $(V,d)$, root $r\in V$ and $g$ groups of 
vertices $\{X_i\sse V\}_{i=1}^g$, and the goal is to find an $r$-tour of minimum length that contains at least one vertex from each group $\{X_i\}_{i=1}^g$.  Our algorithms for the above stochastic problems rely on solving some variants of group Steiner tree. 

\begin{definition}[Group Steiner Orienteering]\label{def:gso}
The input is a metric $(V,d)$, root $r\in V$, $g$ groups of 
vertices $\{X_i\sse V\}_{i=1}^g$ with associated profits $\{\profit_i\}_{i=1}^g$ and a length bound $B$. The goal in \gso is
to compute an $r$-tour of length at most $B$ that maximizes the total profit of covered groups. A group $i\in[g]$
is covered if any vertex from $X_i$ is visited by the tour. 
\end{definition}
\smallskip
An algorithm for \gso is said to be a  $(\beta,\gamma)$-bicriteria approximation algorithm if on any instance of the problem, it finds an $r$-tour of length at most $\gamma\cdot B$ that has profit at least $\frac{1}{\beta}$ times the optimal (which has length at most $B$). 

\smallskip

\begin{definition}[Partial Latency Group Steiner]\label{def:lpgs} The input is 
a metric $(V,d)$, $g$ groups of vertices $\{X_i\sse V\}_{i=1}^g$ with associated weights
$\{w_i\}_{i=1}^g$, root $r\in V$ and a target $h\le g$. The goal in \lpgst is to compute an $r$-tour  $\tau$ 
that covers at least $h$ groups and minimizes the weighted sum of arrival times over all groups. The \emph{arrival  time} of group $i\in[g]$ is the length of the shortest prefix of tour   $\tau$ 
that contains an $X_i$-vertex; if the group is not covered, its arrival time is set to be the entire tour-length. The \lpgst objective  is termed {\em latency}, i.e. 
 \begin{equation}\label{eq:lpgs-obj}
  \ts \text{latency}(\tau) \quad = \quad \sum_{i \text{ covered}} w_i \cdot
  \text{arrival time}_{\tau}(X_i) \,\,\, +  \,\,\, \sum_{i \text{ uncovered}} w_i \cdot
  \text{length}(\tau).
\end{equation}
\end{definition}
\smallskip
An algorithm for \lpgst is said to be a  $(\rho,\sigma)$-bicriteria approximation algorithm if on any instance of the problem, it finds an $r$-tour that covers at least $h/\sigma$ groups and has latency at most $\rho$ times the optimal (which  covers at least $h$ groups). The reason we focus on a bicriteria approximation for \lpgst is that it is harder to approximate than the group Steiner tree problem (see below) and we can obtain  a better bicriteria guarantee for \lpgst. 

To see that \lpgst is at least as hard to approximate as the group Steiner tree problem, consider an arbitrary instance of group Steiner tree with metric $(V,d)$, root $r\in V$ and $g$ groups  $\{X_i\sse V\}_{i=1}^g$. Construct an instance of \lpgst as follows. The vertices are $V'=V\cup\{u\}$ where $u$ is a new vertex. Let $L:=n^2\cdot \max_{a,b} d(a,b)$. The distances in metric $(V',d')$ are: $d'(a,b)=d(a,b)$ if $a,b\in V$ and $d'(a,u)=L+d(a,r)$ if $a\in V$. 
There are $g'=g+1$ groups with $X'_i=X_i$ for $i\in [g]$ and $X'_{g+1}= \{u\}$. The target $h=g$. The weights are $w_i=0$ for $i\in[g]$ and $w_{g+1}=1$. Since the distance from $r$ to $u$ is very large, no approximately optimal \lpgst solution will visit $u$. So any such \lpgst solution covers {\em all} the groups $\{X'_i\}_{i=1}^g$ and has latency equal to the length of the solution (as group $X'_{g+1}$ has weight one  and all others have weight zero).  This reduction also shows that \lpgst on weighted star-metrics (which is used in the \odt algorithm) is at least as hard to approximate as set cover: this is because when  metric $(V,d)$ is a star-metric with center $r$, so is the  new metric $(V',d')$.\footnote{Recall that group Steiner tree on star-metrics is equivalent to the set cover problem.}

\section{Approximation Algorithm for the Isolation Problem} \label{sec:iso}

Recall that an instance of \isoprob is specified by a metric $(V,d)$, a root vertex $r\in V$, and $m$ scenarios
$\{S_i\}_{i=1}^m$ with associated probability values $\{p_i\}_{i=1}^m$. The main result of this section is:
\begin{theorem}\label{thm:isolation}
If there is a $(4,\gamma)$-bicriteria approximation algorithm for group Steiner orienteering then there is an $O(\gamma\cdot \log m)$-approximation algorithm for the isolation problem.
\end{theorem}

We prove this in two steps. First, in Subsection~\ref{subsec:iso-alg} we show that a $(\rho,4)$-bicriteria approximation algorithm for \lpgst can be used to obtain an $O(\rho\cdot \log m)$-approximation algorithm for \isoprob. Then, in Subsection~\ref{subsec:grp-lat} we show that any $(4,\gamma)$-bicriteria approximation algorithm for \gso leads to an $(O(\gamma),4)$-bicriteria approximation algorithm for \lpgst. 

\paragraph{Note on reading this section:} While the results of this section apply to the isolation problem on general metrics, 
readers interested in just the optimal decision tree problem need to only consider weighted star metrics (as discussed after Definition~\ref{def:odt}). In the \odt case, we have the following simplifications (1) a tour   is simply a sequence of tests, (2) the tour length is the sum of test costs in the  sequence, and (3) concatenating tours corresponds to concatenating test sequences. 

\subsection{Algorithm for \isoprob using \lpgst} \label{subsec:iso-alg}
Recall the definition of \isoprob and \lpgst from Section~\ref{sec:prelim}.  Here we will prove:
\begin{theorem}\label{thm:lpg-to-iso}
If there is a $(\rho,4)$-bicriteria approximation algorithm for \lpgst then there is an $O(\rho\cdot \log m)$-approximation algorithm for \isoprob.
\end{theorem}

We first give a high-level description of our algorithm. The algorithm uses an iterative 
approach and maintains a
candidate set of scenarios that contains the realized scenario. In each iteration, the algorithm eliminates a constant fraction of scenarios from the candidate set. So the number of iterations will be bounded by $O(\log m)$. In each iteration we solve a suitable instance of \lpgst in order to refine the candidate set of scenarios.

\paragraph{Single iteration of \isoprob algorithm} As mentioned above, we use \lpgst in each iteration of 
the \isoprob algorithm- we now describe how this is done.
 At the start of each iteration, our algorithm maintains a candidate set $M\sse [m]$ of scenarios that contains the realized scenario.
The probabilities associated with the scenarios $i\in M$ are not the original $p_i$s but their conditional probabilities $q_i:=\frac{p_i}{\sum_{j\in M} p_j}$. The algorithm \palg (given as Algorithm~\ref{alg:palg}) uses \lpgst to compute an $r$-tour $\tau$ such that after
observing the demands on $\tau$, the number of scenarios consistent with these observations is guaranteed to be  a constant factor smaller than $|M|$.

To get some intuition for this algorithm, consider the simplistic case when there is a vertex $u\in V$ located near the root $r$ such that $\approx 50\%$ of the scenarios in $M$ contain it. Then just 
visiting vertex $u$ would reduce the number candidate scenarios by $\approx 50\%$, irrespective of the observation at $u$,
giving us the desired notion of progress. However, each vertex may give a very unbalanced partition of $M$: so we may have to visit multiple vertices before ensuring that the number of candidate scenarios reduces by a constant factor. Moreover, some vertices may
be  too expensive to visit from $r$: so we need to carefully take the metric into account in choosing the set of vertices to visit. Addressing these issues is precisely where the \lpgst problem comes in.

\begin{algorithm}
  \caption{Algorithm $\palg(\;\langle M,\{q_i\}_{i\in M} \rangle\;)$}
  \label{alg:palg}
  \begin{algorithmic}[1]
    \STATE \label{step:palg0} \textbf{let} $g = |M|$. For each $v\in V$,
    define $F_v:=\{i\in M\mid v\in S_i\}$, and
    {\small $D_v := \left\{\begin{array}{ll} F_v & \mbox{ if }|F_v|\le
     g/2\\
    M\setminus F_v &\mbox{ if }|F_v|>  g/2\end{array}\right.$}

    \STATE \label{step:palg-flip}
    \textbf{for each} $i\in M$, set $X_i\leftarrow \{v\in V\mid i\in
    D_v\}$.

    \STATE\label{step:palg1} \textbf{run} the $(\rho,4)$-bicriteria approximation algorithm for \lpgst on the instance 
  with metric $(V,d)$, root $r$, groups
    $\{X_i\}_{i\in M}$ with weights $\{q_i\}_{i\in M}$, and target
    $h:=g-1$. \\
    ~~~~~\textbf{let} $\tau:=r,v_1,v_2,\cdots,v_{t-1},r$ be the $r$-tour
    returned.

    \STATE \label{step:palg2} \textbf{let} $\{P_k\}_{k=1}^t$ be the
    partition of $M$ where
    {\small $P_k:= \left\{
      \begin{array}{ll}
        D_{v_k}\setminus \left(\cup_{j<k} \, D_{v_j}\right) & \text{if }
        1\le k\le t-1\\
        M\setminus \left(\cup_{j<t} \, D_{v_j}\right) & \text{if }
        k=t\end{array}\right.
    $}
    \STATE \textbf{return} tour $\tau$ and the partition
    $\{P_k\}_{k=1}^t$.
  \end{algorithmic}
\end{algorithm}

Note that the information at any vertex $v$ corresponds to a bi-partition $(F_v, M\setminus F_v)$ of the scenario set
$M$, with scenarios $F_v$ having demand at $v$ and scenarios $M\setminus F_v$ having no demand at $v$. So either the
presence of demand or the absence of demand reduces the number of candidate scenarios by half (and represents progress).
To better handle this asymmetry, Step~\ref{step:palg0} associates vertex $v$ with subset $D_v$ which is the smaller of
$\{F_v, \, M\setminus F_v\}$; this corresponds to the set of scenarios under which just the observation at $v$ suffices to 
reduce the number of candidate scenarios below $|M|/2$ (and represents progress). In Steps~\ref{step:palg-flip} and~\ref{step:palg1}, we view vertex $v$ as covering the
scenarios $D_v$.

\paragraph{The overall algorithm for \isoprob} Here we describe how the different iterations are combined to solve \isoprob. 
The final algorithm \iso (given as Algorithm~\ref{alg:isoalg}) is described in a recursive manner where each ``iteration'' is a new call to \iso. As mentioned earlier, at the start of each iteration, the algorithm maintains a
candidate set $M\sse[m]$ of scenarios such that the realized scenario lies in $M$.  Upon observing demands along the tour
produced by algorithm \palg, 
 a new set $M'\sse M$ containing the realized scenario is 
identified such that the number of candidate scenarios reduces by a constant factor (specifically $|M'|\le \frac78 \cdot|M|$).
Then \iso recurses on scenarios $M'$, which corresponds to the next iteration.  After $O(\log m)$ such iterations the realized scenario would be correctly
identified.

\begin{algorithm}
  \caption{Algorithm $\iso\langle M,\{q_i\}_{i\in M} \rangle$}
  \label{alg:isoalg}
  \begin{algorithmic}[1]
    \STATE If $|M|=1$, return this unique scenario as realized.

    \STATE \label{step:iso2} \textbf{run} $\palg\langle M,\{q_i\}_{i\in M} \rangle$ \\
    ~~~~~\textbf{let} $\tau=(r,v_1,v_2,\cdots,v_{t-1},r)$ be the $r$-tour and
     $\{P_k\}_{k=1}^t$ be the partition of~$M$ returned.

    \STATE \textbf{let} $q'_k:=\sum_{i\in P_k} q_i$ \textbf{for all} $k=1\ldots t$.

    \STATE \label{step:iso3} \textbf{traverse} tour $\tau$ and return
    directly to $r$ after visiting the first (if any) vertex $v_{k^*}$
    (for $k^* \in [t-1]$) that determines that the realized scenario is
    in $P_{k^*} \sse M$. If there is no such vertex until the end of
    the tour $\tau$, then set $k^* \leftarrow t$.

    \STATE  \textbf{run} $\iso\langle P_{k^*},
    \{\frac{q_i}{q'_{k^*}}\}_{i\in P_{k^*}}\rangle$ to isolate the
    realized scenario within the subset $P_{k^*}$.
  \end{algorithmic}
\end{algorithm}

\begin{figure}
\begin{center}
 \includegraphics[scale=0.8]{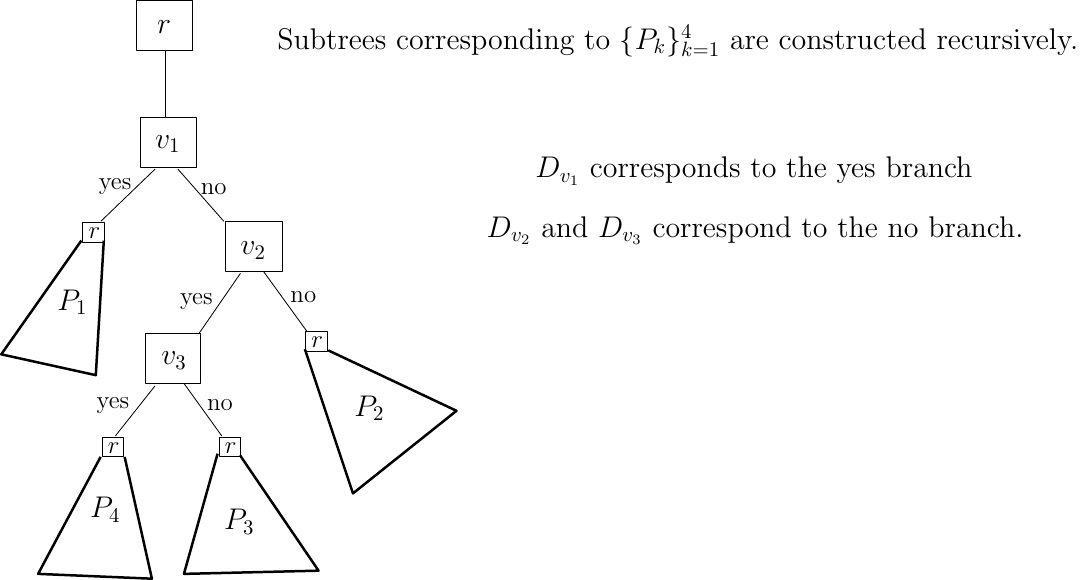}
 \end{center}
\caption{\label{fig:single-phase} Example of decision tree in  single iteration using tour $\tau=(r,v_1,v_2,v_3,r)$.}
\end{figure}

Note that the adaptive Algorithm~\iso implicitly defines a decision tree too: indeed, we create a path $(r, v_1, v_2,
\cdots, v_{t-1}, v_t = r)$, and hang the subtrees created in the recursive call on each instance $\langle P_k,
\{\frac{q_i}{q'_k}\} \rangle$ from the respective node $v_k$. See also Figure~\ref{fig:single-phase}.

\paragraph{Analysis} The rest of this subsection analyzes \iso and proves  Theorem~\ref{thm:lpg-to-iso}.  
We first provide an  outline of the proof. It is easy to show that \iso correctly identifies the realized
scenario after $O(\log m)$ iterations: this is shown formally in Claim~\ref{cl:alg4}. We relate the objective values of 
the \lpgst and \isoprob instances in two steps: Claim~\ref{cl:alg1} shows that \lpgst has a smaller optimal value than \isoprob,  
and Claim~\ref{cl:alg2} shows that any approximate \lpgst solution can be used to construct a {\em partial} \isoprob solution incurring the same cost (in expectation). Since different iterations of \iso deal with different sub-instances of \isoprob, we need to relate the optimal cost of these sub-instances to that of the original instance: this is done in Claim~\ref{cl:alg3}.

Recall that the original instance of \isoprob is defined on metric $(V,d)$, root $r$ and set $\{S_i\}_{i=1}^m$ of scenarios with probabilities $\{p_i\}_{i=1}^m$. \iso works with many sub-instances of the isolation problem. Such an instance $\js$ is specified by a subset $M\sse[m]$ which implicitly defines (conditional) probabilities $q_i=\frac{p_i}{\sum_{j\in M}p_j}$ for all $i\in M$. In other words, $\js$ involves identifying the realized scenario {\em conditioned on} it being in set $M$ (the metric and root remain the same as the original instance). Let $\isoopt(\js)$ denote the optimal value of any instance \js.

\begin{claim}\label{cl:alg1}
  For any instance $\js = \langle M, \{q_i\}_{i\in M}\rangle$, the
  optimal value of the \lpgst instance considered in
  Step~\ref{step:palg1} of algorithm $\palg(\js)$ is at most
  $\isoopt(\js)$.
   \end{claim}
\begin{pf}
  Let $T$ be an optimal decision tree corresponding to  \isoprob
  instance $\js$, and hence $\isoopt(\js) = \isotime(T)$. Note that by
  definition of the sets $\{F_v\}_{v\in V}$, any internal node in
  $T$ labeled vertex $v$ has its two children $v_{yes}$ and $v_{no}$
  corresponding to the realized scenario being in $F_v$ and $M
  \setminus F_v$ (respectively); and by
  definition of $\{D_v\}_{v\in V}$, nodes $v_{yes}$ and $v_{no}$
  correspond  to the realized scenario being in $D_v$ and $M\setminus
  D_v$ (now not necessarily in that order).

We now define an $r$-tour $\sigma$ based on a specific root-leaf path in $T$.   Consider the root-leaf path that at any 
node labeled $v$, moves  to the child $v_{yes}$ or $v_{no}$ that corresponds to
  $M\setminus D_v$, until it reaches a leaf-node $\ell$. Let $r, u_1, u_2, \cdots,
  u_{j}$ denote the sequence of vertices in this root-leaf path, and define $r$-tour   
   $\sigma=\langle r, u_1, u_2, \cdots,
  u_{j}, r\rangle$.  
  Since $T$ is a feasible decision tree for the isolation
  instance,  there is at most one scenario $a\in M$ such that the path $T_{S_a}$ traced in $T$ under demands $S_a$ ends at leaf-node $\ell$. In other words,  every scenario $b\in M\setminus \{a\}$ gives rise to a
  root-leaf path $T_{S_b}$ that diverges from the root-$\ell$ path.  By our definition of the root-$\ell$ path,  the scenarios that diverge  from it are precisely $\cup_{k=1}^{j} D_{u_k}$, and so 
  $\cup_{k=1}^{j} D_{u_k} = M\setminus \{a\}$.
  
  Next, we show that $\sigma$ is a feasible  solution to the
  \lpgst instance in Step~\ref{step:palg1}. By definition of the groups $\{X_i\}_{i\in M}$ (Step~\ref{step:palg-flip} of Algorithm~\ref{alg:palg}), it follows that tour $\sigma$ covers  groups $\cup_{k=1}^{j} D_{u_k}$. So the number of groups covered is at least $|M|-1=h$, and $\sigma$ is a feasible \lpgst solution. 
  
Finally, we bound the \lpgst objective value of $\sigma$ in terms of the isolation cost $\isotime(T)$. To reduce notation let $u_0=r$ below. 
The arrival times in tour $\sigma$ are: 
$$ \mbox{arrival time}_\sigma(X_i) = \left\{
\begin{array}{ll}
\sum_{s=1}^k d(u_{s-1},u_s) & \mbox{ if }i \in D_{u_k} \setminus \cup_{s=1}^{k-1} D_{u_s},\, \mbox{ for }k=1,\cdots,j\\
\mbox{length}(\sigma) & \mbox{ if }i=a
\end{array}
\right.$$
Fix any $k=1,\cdots,j$. For any scenario $i\in D_{u_k}\setminus \cup_{s=1}^{k-1} D_{u_s}$,  the path $T_{S_i}$ traced in $T$ contains the prefix labeled $r,u_1,\cdots,u_k$ of the root-$\ell$ path; so $d(T_{S_i})\ge \sum_{s=1}^k d(u_{s-1},u_s) = \mbox{arrival time}_\sigma(X_i)$. Moreover, for scenario $a$ which is the only scenario not in $\cup_{k=1}^{j} D_{u_k}$,  we have $d(T_{S_a})=\mbox{length}(\sigma) = \mbox{arrival time}_\sigma(X_i)$. Now by~\eqref{eq:lpgs-obj}, $\mbox{latency}(\sigma)\le \sum_{i\in M} q_i\cdot d(T_{S_i})=\isotime(T)=\isoopt(\js)$. 
\end{pf}

\medskip\noindent If we use a $(\rho, 4)$-bicriteria approximation algorithm for \lpgst,
we get the following claim:
\begin{claim}
  \label{cl:alg4}
  For any instance $\js = \langle M, \{q_i\}_{i\in M}\rangle$, the
 latency of tour $\tau$ returned by Algorithm \palg is at most
  $\rho \cdot \isoopt(\js)$.
Furthermore, the resulting partition $\{P_k\}_{k=1}^t$ has each $|P_k| \leq \frac78
  |M|$ for each $k \in [t]$, when $|M|\ge 2$.
\end{claim}
\begin{pf} By Claim~\ref{cl:alg1}, the optimal value of the \lpgst instance in Step~\ref{step:palg1} of algorithm \palg is at
most $\isoopt(\js)$; now the $(\rho,4)$-bicriteria approximation guarantee 
 implies that the latency of the solution tour $\tau$ is at most $\rho$ times that. This proves the first part
of the claim.

Consider $\tau:= \langle r =v_0,v_1,\cdots,v_{t-1},v_t=r\rangle$ the tour returned by the \lpgst algorithm in
Step~\ref{step:palg1} of algorithm \palg; and $\{P_k\}_{k=1}^t$ the resulting partition. 
The $(\rho,4)$-bicriteria approximation guarantee implies  that the number of groups covered by $\tau$ is  $|\cup_{k=1}^{t-1} D_{v_k}|\ge \frac{h}4=
\frac{|M|-1}4\ge \frac{|M|}8$ (when $|M|\ge 2$). By definition of the sets $D_v$, it holds that $|D_v| \le |M|/2$ for
all $v\in V$. Since all but the last part $P_t$ is a  subset of some $D_v$, it holds that $|P_k|\le\frac{|M|}2$ for
$1\le k\le t-1$.  Moreover, the set $P_t$ has size $|P_t|=|M\setminus  (\cup_{j<t} D_{v_j})|\le \frac78 |M|$. This proves the second part of the claim.
\end{pf}

Of course, we don't really care about the latency of the tour \emph{per se}, we care about the expected cost incurred in
isolating the realized scenario. But the two are related (by their very construction), as the following claim
formalizes:

\begin{claim}\label{cl:alg2}
  At the end of Step~\ref{step:iso3} of $\iso\langle M, \{q_i\}_{i\in
    M}\rangle$, the realized scenario lies in $P_{k^*}$. The expected
  distance traversed in this step is at most $2\rho\cdot \isoopt(\langle M, \{q_i\}_{i\in M}\rangle)$.
\end{claim}
\begin{pf}
  Consider the tour $\tau:= \langle r =v_0,v_1,\cdots,v_{t-1},v_t=r\rangle$ returned by
  the \palg algorithm. Recall that visiting any vertex $v$ reveals whether the
  scenario lies in $D_{v}$, or in $M\setminus D_{v}$. In step~\ref{step:iso3} of algorithm \iso,
  we traverse $\tau$ and one of the following happens:
\begin{itemize}
   \item $1\le k^*\le t-1$. Tour returns directly to $r$ from the first vertex $v_k$ (for $1\le
  k\le t-1$) such that the realized scenario lies in $D_{v_k}$; here $k=k^*$. Since the scenario did not lie in any earlier $D_{v_j}$ for $j <  k$, the definition of $P_k = D_{v_k} \setminus (\cup_{j < k} D_{v_j})$
  gives us that the realized scenario is indeed in $P_k$.
  \item $k^*=t$. Tour $\tau$ is completely traversed and we return to $r$. In this case, the
  realized scenario does not lie in any of $\{D_{v_k}\mid 1\le k\le
  t-1\}$, and  it is inferred to be in the complement set $M\setminus (\cup_{j<t} D_{v_j})$,
  which is $P_t$ by definition.
\end{itemize}
Hence for $k^*$ as defined in  Step~\ref{step:iso3} of $\iso\langle M, \{q_i\}_{i\in M}\rangle$, it follows that
$P_{k^*}$ contains the realized scenario; this proves the first part of the claim (and correctness of the algorithm).

For each $i\in M$, let $\alpha_i$ denote the arrival time of group $X_i$ in tour $\tau$; recall that this is the
length of the shortest prefix of $\tau$ until it visits an $X_i$-vertex, and is set to the entire tour length if $\tau$
does not cover $X_i$. The construction of partition $\{P_k\}_{k=1}^t$ from $\tau$ implies that
  \begin{equation*}
    \label{eq:4}
    \ts \alpha_i \,\, = \,\, \sum_{j=1}^k d(v_{j-1},v_j); \qquad \forall i\in
    P_k,~\forall 1\le k\le t,
  \end{equation*}
and hence $\latency(\tau) =\sum_{i\in M} q_i\cdot \alpha_i$.

To bound the expected distance traversed, note the probability that  the traversal returns to $r$ from vertex $v_k$
(for $1\le k\le t-1$)  is exactly $\sum_{i\in P_k} q_i$; with the remaining $\sum_{i\in P_t} q_i$ probability the
entire tour $\tau$ is traversed. Now, using symmetry and triangle-inequality of the distance function $d$, we have $d(v_k,r)\le \sum_{j=1}^k   d(v_{j-1},v_j)$ for all $k\in[t]$. Hence the  expected length traversed is at most:
$$\sum_{k=1}^t \left(\sum_{i\in P_k} q_i \right) \cdot \left(d(v_k,r)+\sum_{j=1}^k
  d(v_{j-1},v_j)\right) \,\,\le \,\,
  2\cdot \sum_{k=1}^t \left(\sum_{i\in P_k} q_i \right) \cdot \left(\sum_{j=1}^k
  d(v_{j-1},v_j)\right)  \,\,=  \,\,2\cdot \sum_{i\in M} q_i\cdot \alpha_i,$$ 
  which is exactly $2\cdot
  \latency(\tau)$.
Finally, by Claim~\ref{cl:alg4}, this is at most $2\cdot \rho\cdot \isoopt(\langle M,
    \{q_i\}_{i\in M}\rangle)$.
\end{pf}

Now, the following simple claim captures the ``sub-additivity'' of $\isoopt$.
\begin{claim}\label{cl:alg3}
  For any instance $\langle M, \{q_i\}_{i\in
    M}\rangle$ and any partition  $\{P_k\}_{k=1}^t$ of $M$,
  \begin{gather}
    \ts \sum_{k=1}^t q'_k \cdot \isoopt(\langle P_k,
    \{\frac{q_i}{q'_k}\}_{i\in P_k}\rangle)\quad \le \quad \isoopt(\langle M,
    \{q_i\}_{i\in M}\rangle),
  \end{gather}
  where $q'_k = \sum_{i\in P_k} q_i$ for all $1\le k\le t$.
\end{claim}
\begin{pf}
Let $T$ denote the optimal decision tree for the instance
  $\js_0:=\langle M, \{q_i\}_{i\in M}\rangle$. For each $k\in [t]$,
  consider instance $\js_k:=\langle P_k, \{\frac{q_i}{q_k'}\}_{i\in
    P_k}\rangle$; a feasible decision tree for instance $\js_k$
  is obtained by taking the decision tree $T$ and considering only paths to
  the leaf-nodes labeled by $\{i\in P_k\}$. Note that this is a feasible
  solution since $T$ isolates all scenarios $\cup_{k=1}^t P_k$.
  Moreover, the expected cost of such a  decision tree for $\js_k$ is $\sum_{i\in
    P_k}\frac{q_i}{q'_k}\cdot d(T_{S_i})$; recall that $T_{S_i}$ denotes the tour traced
    by $T$ under scenario $i\in P_k$. Hence $\opt(\js_k)\le
  \sum_{i\in P_k}\frac{q_i}{q'_k}\cdot d(T_{S_i})$.  Summing over all
  parts $k\in [t]$, we get
  \begin{gather}
    \sum_{k=1}^t q'_k \cdot \opt(\js_k) \quad \le  \quad \sum_{k=1}^t q'_k \cdot
    \sum_{i\in P_k}\frac{q_i}{q'_k}\cdot d(T_{S_i}) \quad = \quad \sum_{i\in M} q_i\cdot
    d(T_{S_i}) \quad = \quad \opt(\js_0),
  \end{gather}
  where the penultimate equality uses the fact that $\{P_k\}_{k=1}^t$ is
  a partition of $M$.
\end{pf}

Given the above claims, we can bound the overall expected cost of the algorithm.
\begin{claim}
  \label{cl:alg4}
  The expected length of the  decision tree  given by $\iso\langle M,
  \{q_i\}_{i\in M}\rangle$ is at most:
  $$2\rho\cdot \log_{8/7} |M|\cdot \isoopt(\langle M, \{q_i\}_{i\in
  M}\rangle).$$
\end{claim}
\begin{pf}
We prove this by induction on $|M|$. The base case of $|M|=1$ is  trivial, since zero length is traversed. Now
consider $|M|\ge  2$. Let instance $\is_0:=\langle M, \{q_i\}_{i\in M}\rangle$. For each $k\in [t]$, consider the
 instance $\is_k:=\langle  P_k, \{\frac{q_i}{q_k'}\}_{i\in P_k}\rangle$, where $q_k' = \sum_{i
    \in P_k} q_i$. Note that $|P_k|\le \frac78 |M|<|M|$ for all $k\in [t]$ by Claim~\ref{cl:alg4} (as $|M|\ge 2$). By the inductive hypothesis, for any $k\in [t]$, the
  expected length of $\iso(\is_k)$ is at most $2\rho\cdot \log_{8/7}
  |P_k|\cdot \isoopt(\is_k)\le 2\rho\cdot (\log_{8/7} |M|-1)\cdot
  \isoopt(\is_k)$, since $|P_k|\le \frac78 |M|$.

  By Claim~\ref{cl:alg2}, the expected length traversed in
  Step~\ref{step:iso3} of $\iso(\is_0)$ is at most $2\rho\cdot
  \isoopt(\is_0)$. The probability of recursing on $\is_k$ is exactly
  $q'_k$ for each $k\in [t]$. So, 
    \begin{eqnarray*}
    \mbox{expected length of }\iso(\is_0) &\le &2\rho\cdot \isoopt(\is_0)+ \ts \sum_{k=1}^t q'_k\cdot \left(\mbox{expected length of }\iso(\is_k)\right)\\
    &\le &2\rho\cdot \isoopt(\is_0)+ \ts \sum_{k=1}^t q'_k\cdot 2\rho \cdot
    (\log_{8/7} |M|-1)\cdot \isoopt(\is_k)\\
    &\le &2\rho\cdot \isoopt(\is_0)+ 2\rho \cdot (\log_{8/7} |M|-1)\cdot
    \isoopt(\is_0)\\
    &= &2\rho \cdot \log_{8/7} |M|\cdot \isoopt(\is_0)
  \end{eqnarray*}
where the third inequality uses Claim~\ref{cl:alg3}.
\end{pf}

Claim~\ref{cl:alg4} implies that our algorithm achieves  an $O(\rho\, \log m)$-approximation 
  for \isoprob. This  completes the proof of Theorem~\ref{thm:lpg-to-iso}.



\subsection{Algorithm for \lpgst using \gso}
\label{subsec:grp-lat}

Recall the definitions of \lpgst and \gso from Section~\ref{sec:prelim}. Here we will prove:
\begin{theorem}\label{thm:lpgs}
If there is a $(4,\gamma)$-bicriteria approximation algorithm for \gso then there is an $(O(\gamma), 4)$-bicriteria approximation algorithm for \lpgst.
\end{theorem}

We now describe the  algorithm for \lpgst in Theorem~\ref{thm:lpgs}.  Consider any instance of \lpgst with metric
$(V,d)$, root $r\in V$, $g$ groups of vertices $\{X_i\sse V\}_{i=1}^g$ having weights $\{w_i\}_{i=1}^g$,  and
target $h\le g$. Let $\zeta^*$ be an optimal tour for the given instance of \lpgst: let $\lat^*$ denote the
latency and $D^*$ the length of $\zeta^*$. We assume (without loss of generality) that the minimum non-zero distance in the metric is one. Let parameter $a:=\frac54$.
 Algorithm~\ref{alg:lpgst} is the
approximation algorithm for \lpgst.
The ``guess'' in the first step means the following. We run the algorithm for all choices of $l$ and return the
solution having minimum latency {\em amongst} those that cover at least $h/4$ groups. Since $1<D^*\le n\cdot \max_e d_e$, the number of choices for $l$ is at most $\log\left(n\cdot \max_{e} d_e\right)$, and so the algorithm runs in polynomial time. 

\begin{algorithm}[ht]
  \caption{Algorithm for \lpgst}
  \label{alg:lpgst}
  \begin{algorithmic}[1]
    \STATE \label{step:grpSt:0} \textbf{guess} an integer $l$ such that $a^{l-1} < D^* \le a^l$.

    \STATE \textbf{mark} all groups as {\em uncovered}.

    \FOR{$i = 1 \ldots l$}

    \STATE \label{step:grpSt:1} \textbf{run} the $(\beta, \gamma)$-bicriteria approximation algorithm for \gso  on the instance with
    groups $\{X_i\}_{i=1}^g$, root $r$, length bound $a^{i+1}$, and profits:
    $$\profit_i:=\left\{
      \begin{array}{ll}
        0 & \mbox{ for each covered group }i\in [g]\\
        w_i & \mbox{ for each uncovered group }i\in [g]
      \end{array}\right.
    $$

    \STATE \textbf{let} $\tau^{(i)}$ denote the $r$-tour obtained above.

    \STATE \textbf{mark} all groups visited by $\tau^{(i)}$ as {\em
      covered}.

    \ENDFOR

    \STATE \label{step:grpSt:3'} \textbf{construct} tour $\tau\leftarrow \tau^{(1)} \circ
    \tau^{(2)} \circ \cdots \circ \tau^{(l)}$, the concatenation of all the above
    $r$-tours. 
    
    \STATE \label{step:grpSt:3} {\em Extend $\tau$ if necessary to ensure that $d(\tau)\ge \gamma\cdot a^l$ (this is only needed for the analysis).}

    \STATE \label{step:grpSt:2} \textbf{run} the $(\beta, \gamma)$-bicriteria approximation algorithm for \gso  on the instance with groups $\{X_i\}_{i=1}^g$, root $r$,
    length bound $a^l$, and {\em unit profit} for each group, i.e. $\profit_i=1$ for all $i\in[g]$. 
    
    \STATE {\bf let} $\sigma$ denote the
  $r$-tour obtained above.

    \STATE \textbf{output} tour $\pi:= \tau\circ\sigma$ as solution to the \lpgst instance.
  \end{algorithmic}
\end{algorithm}

\paragraph{Analysis} 
In order to prove Theorem~\ref{thm:lpgs}, we will show that the algorithm's tour covers at least $\frac{h}4$ groups and has latency $O(\gamma)\cdot \lat^*$.

\begin{claim}\label{cl:lat-gst-1}
  The tour $\tau$ in Step~\ref{step:grpSt:3} has length
  $\Theta(\gamma)\cdot D^*$ and latency $O(\gamma)\cdot \lat^*$.
\end{claim}
\begin{pf}
Due to the $(\beta,\gamma)$-bicriteria approximation guarantee of the \gso algorithm used in Step~\ref{step:grpSt:1}, the length of each $r$-tour $\tau^{(i)}$ is at most $\gamma\cdot a^{i+1}$. So the length of  $\tau$ in Step~\ref{step:grpSt:3'} is at most $\gamma \sum_{i =1}^l a^{i+1} \le \frac{\gamma}{a-1} a^{l+2}  \le  \frac{\gamma a^3}{ a-1} D^*$.
Moreover, the increase in Step~\ref{step:grpSt:3}
  ensures that $d(\tau)\ge \gamma\cdot D^*$. Thus the length of  $\tau$ in Step~\ref{step:grpSt:3'} is $\Theta(\gamma)\cdot D^*$, which proves the first part of the claim.

  The following proof for bounding the latency is based on techniques from
  the {\em minimum latency TSP}~\cite{cgrt,fhr}. Recall the optimal solution $\zeta^*$
  to the \lpgst instance, where $d(\zeta^*)=D^*\in ( a^{l-1},
   a^l]$. For each $i\in [l]$, let $N^*_i$ denote the total weight of
  groups visited in $\zeta^*$ by time $ a^i$; note that $N^*_l$
  equals the total weight of the groups covered by $\zeta^*$.
  Similarly, for each $i\in [l]$, let $N_i$ denote the total weight of
  groups visited in $\tau^{(1)}\cdots \tau^{(i)}$, i.e. by iteration $i$
  of the algorithm. Set $N_0=N^*_0:=0$, and $W:=\sum_{i=1}^g w_i$ the
  total weight of all groups. We have: 
{\small \begin{eqnarray*}
\mbox{latency}(\tau) & \le & \sum_{i=1}^l (N_i-N_{i-1})\cdot \sum_{j=1}^i \gamma a^{j+1} \,\,\, +\,\,\, (W-N_l)\cdot d(\tau) \le  \sum_{i=1}^l (N_i-N_{i-1})\cdot \frac{\gamma a^{i+2}}{ a-1} \,\,\, +\,\,\, (W-N_l)\cdot d(\tau)\\
&= &\sum_{i=1}^l \left( (W-N_{i-1}) - (W-N_i) \right)\cdot \frac{\gamma a^{i+2}}{ a-1} \,\,\, +\,\,\, (W-N_l)\cdot d(\tau) \le  \sum_{i=0}^l (W-N_i) \cdot \frac{\gamma a^{i+3}}{ a-1} \quad =:\quad T.
\end{eqnarray*}  }
  The last inequality uses the bound $d(\tau)\le \frac{\gamma}{ a-1}  a^{l+2}$ from above.
  
The latency of the optimal tour $\zeta^*$ is
\begin{eqnarray*}
\lat^* &\ge & \sum_{i=1}^{l-1}  a^{i-1}(N^*_i-N^*_{i-1})\,\,\, +\,\,\, (W-N^*_l)\cdot D^*\\
&\ge & \sum_{i=1}^{l-1}  a^{i-1}\left((W-N^*_{i-1})-(W-N^*_i)\right)\,\,\, +\,\,\, (W-N^*_l)\cdot  a^{l-1}\quad \ge \quad (1-\frac1 a)\sum_{i=0}^{l}  a^{i}(W-N^*_{i}).
\end{eqnarray*}

  Consider any iteration $i\in [l]$ of the algorithm in
  Step~\ref{step:grpSt:1}. Note that the optimal value of the \gso
  instance solved in this iteration is at least $N^*_i-N_{i-1}$: the
  $ a^i$ length prefix of tour $\zeta^*$ corresponds to a feasible
  solution to this \gso instance with profit at least $N^*_i-N_{i-1}$. The \gso algorithm implies
  that the profit obtained in $\tau^{(i)}$, i.e.  $N_i-N_{i-1}\ge
  \frac14\cdot (N^*_i-N_{i-1})$, i.e. $W-N_i\le \frac34\cdot (W-N_{i-1})
  + \frac14 \cdot (W-N^*_i)$. Using this,
  \begin{eqnarray*}
    ( a-1)\frac{T}{\gamma}  &= &\sum_{i=0}^l  a^{i+3}\cdot (W-N_i) \quad \le \quad  a^3\cdot W + \frac14 \sum_{i=1}^l  a^{i+3} (W-N^*_{i})+ \frac34
    \sum_{i=1}^l  a^{i+3} (W-N_{i-1}) \\
&\le &\frac{ a^4}{ a-1} \cdot \lat^* + \frac34 \sum_{i=1}^l  a^{i+3} (W-N_{i-1}) \quad = \quad \frac{ a^4}{ a-1} \cdot \lat^* + \frac{3 a}4 \sum_{i=0}^{l-1}  a^{i+3} (W-N_{i}) \\
    &\le &\frac{ a^4}{ a-1} \cdot \lat^* + \frac{3 a}4 \cdot ( a-1)\frac{T}\gamma
  \end{eqnarray*}
This implies $T\le \gamma\cdot \frac{ a^4}{( a-1)^2(1-3 a/4)}\cdot \lat^*=O(\gamma)\cdot \lat^*$
since $ a=\frac54$. This completes the proof.
\end{pf}

\begin{claim}\label{cl:lat-gst-2}
  The tour $\sigma$ in Step~\ref{step:grpSt:2} covers at least
  $\frac{h}{4}$ groups and has length $O(\gamma)\cdot D^*$.
\end{claim}
\begin{pf}
  Since we know that the optimal tour $\zeta^*$ has length at most
  $ a^l$ and covers at least $h$ groups, it is a feasible solution to
  the \gso instance defined in Step~\ref{step:grpSt:2}. So the \gso algorithm 
  ensures that the tour $\sigma$ has length at most $\gamma  a^l =
  O(\gamma) D^*$ and profit (i.e. number of groups) at least $h/4$. 
\end{pf}

\begin{lemma}\label{th:grp-lat}
  Tour $\pi = \tau \cdot \sigma$ covers at least $\frac{h}{4}$ groups
  and has latency $O(\gamma)\cdot \lat^*$.
\end{lemma}
\begin{pf}
  Since $\pi$ visits all the vertices in $\sigma$,
  Claim~\ref{cl:lat-gst-2} implies that $\pi$ covers at least
  $\frac{h}4$ groups. For each group $i\in[g]$, let $\alpha_i$ denote
  its \emph{arrival time} under the tour $\tau$ after 
  Step~\ref{step:grpSt:3}---recall that the arrival time $\alpha_i$ for
  any group $i$ that is not covered by $\tau$ is set to the length of
  the tour $d(\tau)$. Claim~\ref{cl:lat-gst-1} implies that the latency
  of tour $\tau$, $\sum_{i=1}^g w_i\cdot \alpha_i=O(\gamma)\cdot \lat^*$.
  Observe that for each group $i$ that is {\em covered} in $\tau$, its arrival
  time under tour $\pi = \tau \cdot \sigma$ remains $\alpha_i$. For any
  group $j$ {\em not covered} in $\tau$, its arrival time under $\tau$
is $d(\tau)\ge \gamma\cdot a^l$ (due to Step~\ref{step:grpSt:3}), and its arrival time under $\pi$ is $d(\pi)\le O(\gamma) \cdot D^* = O(1)\cdot d(\tau)$. Hence, the arrival time under $\pi$ of each group $i\in
  [g]$ is $O(1)\cdot \alpha_i$, i.e., at most a constant factor more
  than its arrival time in $\tau$. Now using Claim~\ref{cl:lat-gst-1}
  completes the proof.
\end{pf}

Finally, Lemma~\ref{th:grp-lat} directly implies Theorem~\ref{thm:lpgs}.

\medskip
\noindent {\bf Remark:}  The above approach also leads to an 
 approximation algorithm for the {\em minimum latency group Steiner}
problem, which is the special case of \lpgst when the target $h=g$. 
\smallskip

\begin{definition}[Minimum Latency Group Steiner]\label{def:lgst} The input is 
a metric $(V,d)$, $g$ groups of vertices $\{X_i\sse V\}_{i=1}^g$ with associated non-negative weights
$\{w_i\}_{i=1}^g$ and root $r\in V$. The goal in \lgs is to compute an $r$-tour 
 that covers all groups with positive weight and minimizes the weighted sum of arrival times of the groups.  
The \emph{arrival  time} of group $i\in[g]$ is the length of the shortest prefix of the tour  
that contains a vertex from $X_i$.
\end{definition}
\smallskip
Note that the objective here is to minimize the sum
of weighted arrival times where  every group has to be visited. 
 The algorithm for latency group Steiner is in fact simpler than Algorithm~\ref{alg:lpgst}: we do not need the ``guess'' $l$ (Step~\ref{step:grpSt:0}) and we just repeat Step~\ref{step:grpSt:1} until {\em all} groups are covered (instead of stopping after $l$
iterations). A proof identical to that in Claim~\ref{cl:lat-gst-1} gives:
\begin{corollary}\label{cor:lgs} 
If there is a $(4,\gamma)$-bicriteria approximation algorithm for \gso then there is an $O(\gamma)$-approximation algorithm for the latency group Steiner problem.
\end{corollary}

Combined with the $(4,O(\log^2 n)$-bicriteria approximation algorithm for \gso (see Section~\ref{subsec:gso}) we obtain an $O(\log^2n)$-approximation algorithm for \lgs.  It is  shown in~\cite{Vish-thesis} that any $\alpha$-approximation algorithm for \lgs can be used to obtain an $O(\alpha\cdot\log g)$-approximation algorithm for group Steiner tree. Thus improving this
$O(\log^2n)$-approximation algorithm for latency group Steiner would also improve the best known bound for the standard group
Steiner tree problem.


\section{Optimal Decision Tree Problem}
\label{sec:odt}

Recall that  the \emph{\dtp} consists of  a set of diseases with their probabilities (where exactly one disease occurs) 
and a set of binary tests with costs, and the goal is to identify the realized disease at minimum expected cost. In this section we prove Theorem~\ref{thm:main1}.

\begin{figure}[h]
\begin{center}
\includegraphics[scale=0.68]{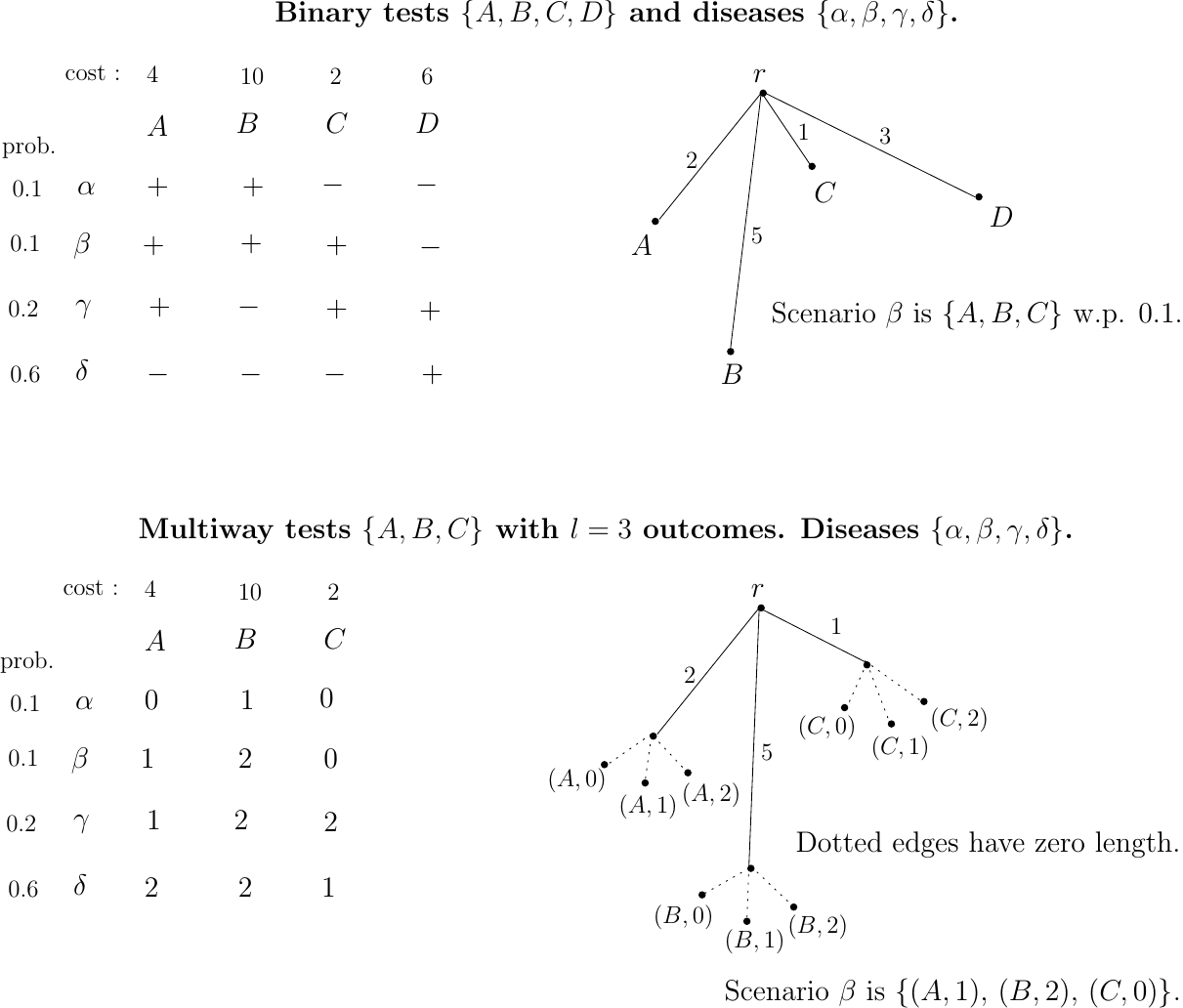}
\end{center}
\caption{\label{fig:odt-redn} Reducing optimal decision tree to Isolation: binary tests (top), multiway tests
(bottom).}
\end{figure}

As noted in Section~\ref{sec:prelim} the \dtp (Definition~\ref{def:odt})  is a special
  case of \isoprob (Definition~\ref{def:iso}). We recall the reduction for convenience. Given an instance of \odt, consider a metric $(V,d)$ induced by a weighted star with
  center $r$ and $n$ leaves corresponding to the tests. For each $j\in
  [n]$, we set $d(r,j)=\frac{c_j}2$. The demand scenarios are as
  follows: for each $i\in [m]$ scenario $i$ has demands $S_i=\{j\in [n]\mid i\in T_j\}$.
  It is easy to see that this \isoprob instance corresponds
  exactly to the optimal decision tree instance. Figure~\ref{fig:odt-redn} gives an example.

The main observation here is the following:  

\begin{theorem}
  \label{th:gso-stars} There is a $(1-\frac1e)$-approximation algorithm for the group Steiner
  orienteering problem on weighted star metrics.
\end{theorem}
\begin{pf}
Consider an instance of \gso (Definition~\ref{def:gso}) on  weighted star-metric
  $(V,d)$ with center $r$ (which is also the root in \gso) and leaves $[n]$, $g$ groups $\{X_i\sse
  [n]\}_{i=1}^g$ with profits $\{\profit_i\}_{i=1}^g$, and length bound $B$.
  If for each $j\in[n]$, we define set $Y_j:= \{i\in [g]\mid j\in X_i\}$ of
  cost $c_j:=\frac{d(r,j)}2$, then solving the \gso instance is the same
  as computing a collection $K\sse [n]$ of the sets with $\sum_{j\in K}
  c_j\le B/2$ that maximizes $f(K):= \sum\{ \profit_i\mid i\in \cup_{j\in K}
  Y_j\}$. But the latter problem is precisely an instance of maximizing a monotone submodular
  function over a knapsack constraint ($\sum_{j\in K}
  c_j\le B/2$), for which a $(1-\frac1e)$-approximation algorithm is
  known~\cite{s04}.
\end{pf}

\medskip Combining this result with Theorem~\ref{thm:isolation}, we obtain an $O(\log m)$ approximation algorithm for \isoprob on weighted star-metrics and hence \odt. This
proves the first part of Theorem~\ref{thm:main1}.

\paragraph{Multiway tests.}  
Our algorithm
can be easily extended to the generalization of \odt where tests have multiway (instead of binary) outcomes.  In this setting (when each test has at
most $l$ outcomes), any test $j\in [n]$ induces a partition $\{T_j^k\}_{k=1}^l$ of $[m]$ into $l$ parts (some of them may be empty), and performing test $j$
determines which part the realized disease lies in. Note that this problem is also a special case of
\isoprob. As before, consider a metric $(V,d)$ induced by a weighted star with center $r$ and $n$ leaves corresponding
to the tests. For each $j\in [n]$, we set $d(r,j)=\frac{c_j}2$. Additionally, for each $j\in [n]$, introduce $l$ copies
of test-vertex $j$, labeled $(j,1),\cdots,(j,l)$, at zero distance from each other. The demand scenarios are defined
naturally: for each $i\in [m]$, scenario $i$ has demands $S_i=\{(j,k)\mid i\in T^k_j\}$. See also an example in
Figure~\ref{fig:odt-redn}. Clearly this \isoprob instance is equivalent to the (multiway) decision tree instance.
Since the resulting metric is still a weighted star (we only made vertex copies), Theorem~\ref{th:gso-stars} along with Theorem~\ref{thm:isolation} implies an $O(\log m)$-approximation for the multiway decision tree
problem. This proves the second part of Theorem~\ref{thm:main1}.


\section{Adaptive Traveling Salesman Problem}\label{sec:stsp}

Recall that the adaptive TSP (Definition~\ref{def:stsp}) 
consists of a metric $(V,d)$ with root $r\in V$ and demand distribution $\ds$, and the goal  is to visit all demand vertices (drawn from $\ds$) using an $r$-tour of minimum  expected cost.   We first show the following simple fact relating this problem to the isolation problem.

\begin{lemma}
  \label{lem:reduce-to-isol}
If there is an $\alpha$-approximation algorithm for \isoprob then there is an
  $\left(\alpha+\frac32\right)$-approximation algorithm for \stsp.
\end{lemma}
\begin{pf}
We first claim that any feasible solution $T$ to \stsp is also feasible for \isoprob. For this it suffices to show that
the paths $T_{S_i}\ne T_{S_j}$ for any two scenarios $i,j\in[m]$ with $i\ne j$. Suppose (for a contradiction) that paths
$T_{S_i}=T_{S_j}=\pi$ for some $i\ne j$. By feasibility of $T$ for \stsp, path $\pi$ contains all vertices in
$S_i\bigcup S_j$. Since $S_i\ne S_j$, there is some vertex in $(S_i\setminus S_j)\bigcup (S_j\setminus S_i)$; let $u\in
S_i\setminus S_j$ (the other case is identical). Consider the point where $\pi$ is at a node labeled $u$: then path
$T_{S_i}$ must take the $yes$ child, whereas path $T_{S_j}$ must take the $no$ child. This contradicts the assumption
$T_{S_i}=T_{S_j}=\pi$. Thus any solution to \stsp is also feasible for \isoprob; moreover the expected cost remains the
same. Hence the optimal value of \isoprob is at most that of \stsp.

Now, using any $\alpha$-approximation algorithm for \isoprob, we obtain a decision tree $T'$ that isolates the realized scenario and
has expected cost $\alpha \cdot \opt$,  where $\opt$ denotes the optimal value of the \stsp instance. This suggests the
following feasible solution for \stsp:
\begin{enumerate}
 \item Implement $T'$ to determine the realized scenario $k\in [m]$, and return to $r$.
 \item Traverse a $\frac32$-approximate TSP tour~\cite{c} on vertices $\{r\}\bigcup S_k$.
\end{enumerate}
From the preceding argument, the expected length in the first phase is at most $\alpha\cdot \opt$. The expected length
in the second phase is at most $\frac32 \sum_{i=1}^m p_i\cdot \tsp(S_i)$, where $\tsp(S_i)$ denotes the minimum length
of a TSP tour on $\{r\}\bigcup S_i$. Note that $\sum_{i=1}^m p_i\cdot \tsp(S_i)$ is a lower bound on the optimal \stsp
value. So we obtain a solution that has expected cost at most $(\alpha + \frac32) \opt$, as claimed.
\end{pf}

\medskip

Therefore, it suffices to obtain an approximation algorithm for \isoprob. In the next subsection we obtain a $(4,O(\log^2n))$-bicriteria approximation  algorithm for \gso, which combined with Theorem~\ref{thm:isolation} and Lemma~\ref{lem:reduce-to-isol} yields an $O(\log^2n\cdot \log m)$-approximation algorithm for both \isoprob and \stsp.  This would prove Theorem~\ref{thm:main2}.  

\subsection{Algorithm for Group Steiner Orienteering} \label{subsec:gso}
Recall the \gso problem (Definition~\ref{def:gso}). 
Here  we obtain a bicriteria approximation algorithm for \gso.
\begin{theorem}
  \label{thm:gso}
  There is a $(4, O(\log^2n))$-bicriteria approximation algorithm for
  \gso, where $n$ is the number of vertices in the metric. That is, the algorithm's tour has length $O(\log^2n)\cdot B$ and has profit at least $\frac{1}{4}$ times the optimal profit of a length $B$ tour.
\end{theorem}

This algorithm is based on a 
greedy framework that is used in many maximum-coverage problems: the solution is constructed iteratively where each iteration adds an $r$-tour that maximizes the ratio of profit to length. In order to find an $r$-tour (approximately) maximizing the profit to length ratio, we use a slight modification of an existing algorithm~\cite{ccgg}; see Theorem~\ref{thm:gso-ratio} below.  The final \gso algorithm is then given as Algorithm~\ref{alg:gsoalg}.

\begin{theorem}  \label{thm:gso-ratio}
There is a polynomial time algorithm that given any instance of \gso, outputs an $r$-tour $\sigma$ having profit-to-length ratio $\frac{\profit(\sigma)}{d(\sigma)} \ge \frac{1}{\alpha}\cdot\frac{\opt}{B}$. Here  $\profit(\sigma)$ and $d(\sigma)$ denote the profit and length (respectively) of tour $\sigma$, \opt is the optimal value of the \gso instance, $B$ is the length bound in \gso and $\alpha=O(\log^2n)$ where $n$ is the number of vertices in the metric.
\end{theorem}
\begin{pf} This result essentially follows from~\cite{ccgg}, but requires some modifications  which we present here for completeness. 
We first preprocess the metric to only include
  vertices within distance $B/2$ from the root $r$: note that since the
  optimal \gso tour cannot visit any excluded vertex, the optimal profit
  remains unchanged by this. To reduce notation, we refer to this restricted vertex-set also as $V$ and let $|V|=n$.
  We denote the set of all edges in the metric by $E={V \choose 2}$.
We assume (without loss of generality) that every group is covered by some vertex in $V$; otherwise the group can be dropped from the \gso instance. By averaging, there  is some vertex $u\in V$ covering groups of total profit at least $\frac1n \sum_{i=1}^g \profit_i$. If $\opt\le \frac4{n} \sum_{i=1}^g \profit_i$ then the $r$-tour that just visits vertex $u$ has profit-to-length ratio at least $\frac{\opt}{4B}$ and is output as the desired tour $\sigma$. Below we assume that $\frac1n \sum_{i=1}^g \profit_i < \frac{\opt}{4}$. 

\def\lpgo{\ensuremath{\mathsf{LP}_\gso}\xspace}

We use the following linear programming relaxation \lpgo for \gso:
      \begin{eqnarray}
        \max & \sum_{i=1}^g  \profit_i\cdot y_i & 
         \label{lp:gso}\\
        \mbox{s.t.}& x(\delta(S))\ge y_i& \quad \forall S\sse V : r\not\in S,~X_i\sse S;~~\forall i\in[g] \notag \\
        &\sum_{e\in E} d_e\cdot x_e \le B&\notag \\
        & 0\le y_i\le 1 & \quad \forall i\in [g]\notag \\
        & x_e\ge 0& \quad \forall e\in E\notag 
      \end{eqnarray}
It is easy to see that this a valid relaxation of \gso: any feasible \gso solution corresponds to a feasible solution above where the $x,y$ variables are $\{0,1\}$ valued. So the optimal value $\sum_{i=1}^g  \profit_i\cdot y_i \ge \opt$. The algorithm is given as Algorithm~\ref{alg:densityGSO} and uses the following known results: Theorem~\ref{thm:ccgg} shows how to round fractional solutions to \lpgo on tree metrics and Theorem~\ref{thm:frt} shows how to transform an \lpgo solution on general metrics to one on a tree. 
\begin{theorem}[\cite{ccgg}]\label{thm:ccgg}
There is a polynomial time algorithm that given any fractional solution $(x,y)$ to \lpgo on a tree metric where all variables are integral multiples of $\frac{1}{N}$, finds a subtree $A$ containing $r$ such that $\frac{d(A)}{\profit(A)}\le O(\log N)\cdot \frac{\sum_{e\in E} d_e\cdot x_e}{\sum_{i=1}^g \profit_i\cdot y_i}$. Here  $\profit(A)$ and $d(A)$ denote the profit and length (respectively) of subtree $A$.
\end{theorem}

\begin{theorem}[\cite{FRT03}]\label{thm:frt} There is a polynomial time algorithm that given any metric $(V,d)$ with edges $E={V\choose 2}$ and capacity function $x:E\rightarrow\mathbb{R}_+$, computes a spanning tree $T$ in this metric such that $\sum_{f\in T} d_f\cdot x_T(f) \le O(\log n)\cdot \sum_{e\in E} d_e\cdot x(e)$, where
$$x_T(f)\quad :=\quad \sum_{u,v \,:\, f\in uv\, path\, in\, T}\,\,\, x(u,v),\qquad \forall f\in T.$$
\end{theorem}

\begin{algorithm}[!h]
  \caption{Algorithm for \gso maximizing profit-to-length ratio.}
\label{alg:densityGSO}
\begin{algorithmic}[1]
    \STATE {\bf solve} the linear program \lpgo to obtain solution $(x,y)$.
    \STATE {\bf run} the algorithm from Theorem~\ref{thm:frt} on metric $(V,d)$ with edge-capacities $x$ to obtain a spanning tree $T$ with ``new capacities'' $x_T$ on edges of $T$. 
 \STATE \label{step:gso-round} {\bf round} down each $x_T(e)$ to an integral multiple of $\frac1{n^3}$.
    \STATE \label{step:gso-flow} {\bf for} each group $i\in[g]$, let $y'_i$ be
      the maximum flow from $r$ to group $X_i$ under  
      capacities $x_T$.
      
    \STATE \label{step:gso-density} {\bf run} the algorithm from Theorem~\ref{thm:ccgg} using variables $x_T$ and $y'$ to obtain subtree $A$. 
    \STATE {\bf output} an Euler tour $\sigma$ of the subtree $A$.
  \end{algorithmic}
\end{algorithm}

By definition of the new edge-capacities $x_T$ on edges of $T$ (see Theorem~\ref{thm:frt}) it is clear that the capacity of each cut under $x_T$ is at least as much as under $x$; i.e. $\sum_{e\in \delta(S)} x_T(e)\ge 
\sum_{e\in \delta(S)} x(e)$ for all $S\sse V$.
For each group $i\in [g]$, since  capacities $x$ support $y_i$ units of flow from $r$ to $X_i$, it follows that the new capacities $x_T$ on tree $T$ also support such a flow. So $(x_T,y)$ is a feasible solution to \lpgo on tree $T$ with budget $O(\log n)\cdot B$.  
In order to apply the rounding algorithm from~\cite{ccgg} for \gso on trees, we need to ensure the technical condition (see Theorem~\ref{thm:ccgg}) that every variable is an integral multiple of $\frac1N$ for some $N=poly(n)$. This is the reason behind modifying capacities $x_T$ in Step~\ref{step:gso-round}. Note that this step reduces the capacity $x_T(e)$ of each edge $e\in T$ by at most   $\frac1{n^3}$. Since any cut in tree $T$ has at most $n$ edges,
  the capacity of any cut decreases by at most $\frac1{n^2}$ after
  Step~\ref{step:gso-round}; and by the max-flow min-cut theorem,
  the maximum flow value for group $X_i$ is $y'_i\ge y_i-\frac1{n^2}$ for each $i\in [g]$ (in
  Step~\ref{step:gso-flow}). Furthermore, since all edge capacities are
  integer multiples of $\frac1{n^3}$, so are all the flow values 
  $y'_i$s.  
  So $(x_T,y')$ is a feasible solution to \lpgo on tree $T$ (with budget $O(\log n)\cdot B$) that satisfies the condition required in Theorem~\ref{thm:ccgg}, with $N=n^3$.  
Also note  that this rounding down does not change the fractional profits much,  since
  \begin{equation}
    \label{eq:gso-ratio1}
    \sum_{i=1}^g  \profit_i\cdot y'_i \,\, \ge   \,\, \sum_{i=1}^g
     \profit_i\cdot y_i - \frac{1}{n^2}\sum_{i=1}^g
     \profit_i  \,\,\, \ge  \,\,\, \frac34\cdot \opt - \frac{1}{n^2}\sum_{i=1}^g
    \profit_i  \,\,\, \ge  \,\,\, \frac34\cdot \opt - \frac{\opt}{4n}\,\,\ge \,\,\frac{\opt}{2}
  \end{equation}
where the second last inequality follows from $\frac1n \sum_{i=1}^g  \profit_i \le \frac{\opt}4$ (by the preprocessing). Now applying Theorem~\ref{thm:ccgg} implies that subtree $A$ satisfies:
\begin{eqnarray*}\frac{d(A)}{\profit(A)} &\le_{(Theorem~\ref{thm:ccgg})} & \,\, O(\log N)\cdot \frac{\sum_{e\in T} d_e\cdot x_T(e)}{\sum_{i=1}^g \profit_i\cdot y'_i}\quad \le_{\eqref{eq:gso-ratio1}} \quad O(\log N)\cdot \frac{\sum_{e\in T} d_e\cdot x_T(e)}{\opt}\\
&\le_{(Theorem~\ref{thm:frt})} \,\, & O(\log N\,\log n)\cdot \frac{\sum_{e\in E} d_e\cdot x(e)}{\opt}\quad \le \quad O(\log^2n)\cdot \frac{B}{\opt}.\end{eqnarray*}
Finally, since we output an Euler tour of $A$, the theorem follows.
\end{pf}
\medskip

\noindent
{\bf Remark:} A simpler approach in Theorem~\ref{thm:gso-ratio} might have been  to use the randomized algorithm from~\cite{gkr} rather than the deterministic algorithm (Theorem~\ref{thm:ccgg}) from~\cite{ccgg}. This however does not work directly since \cite{gkr} only yields a random solution $A'$ with  expected length $\E[d(A')] \le O(\log n)\cdot  \sum_{e\in E} d_e\cdot x_e$ and   expected profit $\E[\phi(A')]\ge  \sum_{i=1}^g \profit_i\cdot y_i$. While this does guarantee the existence of a solution with length-to-profit ratio at most $O(\log n)\cdot \frac{\sum_{e\in E} d_e\cdot x_e}{\sum_{i=1}^g \profit_i\cdot y_i}$, it may not find such a solution with reasonable (inverse polynomial) probability. 

\paragraph{Algorithm}
The \gso algorithm first preprocesses the metric to only include  vertices within distance $B/2$ from the root $r$: note that the optimal profit remains unchanged by this.   The algorithm then follows a standard greedy approach (see
  eg. Garg~\cite{garg0}), and is given as Algorithm~\ref{alg:gsoalg}. 
  
\begin{algorithm}[!h]
  \caption{Algorithm for \gso.}
\label{alg:gsoalg}
\begin{algorithmic}[1]
    \STATE {\bf initialize} $r$-tour $\tau\leftarrow \emptyset$ and mark  all groups as uncovered.
    \WHILE{length of $\tau$ does not exceed $\alpha\cdot B$}
    \STATE {\bf set} residual profits:
      $$\widetilde{\profit}_i:=\left\{
        \begin{array}{ll}
          0 & \mbox{ for each covered group }i\in [g]\\
          \profit_i & \mbox{ for each uncovered group }i\in [g]
        \end{array}\right.
      $$
    \STATE\label{step:gso-ratio} {\bf run} the algorithm from Theorem~\ref{thm:gso-ratio} on the \gso instance with profits 
    $ \widetilde{\profit}$ to obtain $r$-tour $\sigma$.

   \STATE \label{step:gso-case1} {\bf if} $d(\sigma)\le \alpha B$ then
      $\tau'\leftarrow \tau\circ \sigma$.
    \STATE \label{step:gso-case2} {\bf if} $d(\sigma)> \alpha B$ then:
\begin{itemize}
\item[(i)] partition tour $\sigma$ into at most $2\cdot \frac{d(\sigma)}{\alpha B}$ paths,
      each of length at most $\alpha B$;
    \item[(ii)]  let $\sigma'$ denote the path containing maximum profit;
    \item[(iii)]  let $\langle r,\sigma',r\rangle$ be the $r$-tour obtained by connecting both end-vertices of path $\sigma'$ to $r$.
    \item[(iv)] set $\tau'\leftarrow \tau\cup \langle r,\sigma',r\rangle$.
    \end{itemize}
    \STATE {\bf set} $\tau\leftarrow \tau'$. Mark all groups visited in $\tau$ as
      covered.
  \ENDWHILE
  \STATE {\bf output} the $r$-tour $\tau$.
  \end{algorithmic}
\end{algorithm}

  \paragraph{Analysis}
  Let $\opt$ denote the optimal profit of the given \gso instance.
 In the following, let
  $\alpha:=O(\log^2n)$ which comes from Theorem~\ref{thm:gso-ratio}.
  We  prove that Algorithm~\ref{alg:gsoalg} achieves a $(4,2\alpha
  +1)$ bicriteria
  approximation guarantee, i.e. solution $\tau$ has profit at least $\opt/4$ and length $(2\alpha
  +1)\cdot B$. 
  
  By the description of the algorithm, we iterate as long
  as the total length of edges in $\tau$ is at most  $\alpha B$. Note that the increase in length of $\tau$ in any iteration is at most
  $(\alpha+1)\cdot B$ since every vertex is at distance at
  most $B/2$ from $r$. So the final length $d(\tau)\le (2\alpha
  +1)\cdot B$. This proves
  the bound on the length. 
  
  It now
  suffices to show that the final subgraph $\tau$ gets profit at least
  $\frac\opt{4}$. At any iteration, let $\profit(\tau)$ denote the
  profit of the current solution $\tau$, and $d(\tau)$ its length. Since $d(\tau)>\alpha B$ upon
  termination, it suffices to show the following invariant 
  over the iterations of the algorithm:
  \begin{equation}
    \label{eq:gso-greedy}
    \profit(\tau) \quad \ge  \quad \min\left\{ \frac{\opt}4, \frac{\opt}{2\alpha B}\cdot
      d(\tau)\right\}
  \end{equation}

At the start of the algorithm, inequality~\eqref{eq:gso-greedy} holds trivially
  since $d(\tau) = 0$ for $\tau=\emptyset$. Consider any iteration where
  $\profit(\tau)<\opt/4$ at the beginning: otherwise~\eqref{eq:gso-greedy} trivially holds for the next iteration. The
  invariant now ensures that $d(\tau) < \alpha B/2$ and hence we proceed
  further with the iteration. Moreover, in Step~\ref{step:gso-ratio} the optimal value
  of the ``residual'' \gso instance with profits $\widetilde \profit$ is $\widetilde\opt \ge \opt - \profit(\tau) \ge \frac34\cdot \opt$ (by considering the optimal tour for the \gso instance with profits $\profit$). By Theorem~\ref{thm:gso-ratio}, the $r$-tour $\sigma$ satisfies $d(\sigma)/\widetilde\profit(\sigma) \le \alpha\cdot B/\widetilde\opt\le 2\alpha\cdot B/\opt$.

We finish by handling the two possible cases
(Steps~\ref{step:gso-case1} and~\ref{step:gso-case2}).
  \begin{itemize}
  \item If $d(\sigma)\le \alpha B$, then $\profit(\tau')=\profit(\tau)+\widetilde\profit(\sigma)\ge \frac{\opt}{2\alpha B}\cdot d(\tau) + \frac{\opt}{2\alpha B}\cdot d(\sigma)=\frac{\opt}{2\alpha B}\cdot d(\tau')$.
  \item If $d(\sigma)> \alpha B$, then $\sigma$ is partitioned into at most $\frac{2d(\sigma)}{\alpha B}$ paths of length $\alpha  B$ each. The path $\sigma'$ of best profit has $\widetilde\profit(\sigma')\ge \frac{\alpha B}{2\cdot d(\sigma)} \widetilde\profit(\sigma)\ge \frac{\opt}4$; so $\profit(\tau')\ge\widetilde\profit(\sigma')\ge  \frac\opt{4}$.
  \end{itemize}
In either case $r$-tour $\tau'$ satisfies  inequality~\eqref{eq:gso-greedy}, and since $\tau \gets \tau'$ at the end of  the
iteration, the invariant holds for next iteration as well. This completes the proof of Theorem~\ref{thm:gso}.

\section{Adaptive Traveling Repairman}
\label{sec:strp}

In this section we consider the adaptive traveling repairman problem (\strp), where given a demand distribution, the goal is to
find an adaptive strategy that minimizes the expected {\em sum  of arrival times} at demand vertices. As in adaptive TSP, we assume that the demand distribution $\ds$ is specified explicitly in terms of its support.

\smallskip 
\begin{definition}[Adaptive Traveling Repairman]
The input is a metric $(V,d)$, root $r$ and demand distribution $\ds$ given by $m$ distinct subsets $\{S_i\}_{i=1}^m$ with  probabilities $\{p_i\}_{i=1}^m$ (which sum to one). The goal in \strp is to compute a decision tree $T$ in
metric $(V,d)$ such that:
  \begin{itemize}
  \item the root of $T$ is labeled with the root vertex $r$, and
  \item for each scenario $i\in[m]$, the path $T_{S_i}$ followed on input $S_i$ contains \underline{all} vertices in $S_{i}$.
  \end{itemize}
  The objective function is to minimize the expected latency
  $\sum_{i=1}^m p_i \cdot \lat(T_{S_i})$, where $\lat(T_{S_i})$ is the sum of arrival times at vertices $S_i$ along path $T_{S_i}$.
\end{definition}

\smallskip

We obtain an $O(\log^2n\,\log m)$-approximation algorithm for \strp (Theorem~\ref{thm:main3}). The high-level approach here is similar to that for \stsp, but there are some important differences. Unlike \stsp, we can not directly reduce \strp to the isolation problem: so there is no analogue of Lemma~\ref{lem:reduce-to-isol} here.  The following example illustrates this.

\smallskip

\begin{example}
Consider an instance of \strp 
on a star-metric with center $r$ and leaves $\{v, u_1,\cdots,u_n\}$. Edges $(r,u_i)$ have unit length for each $i\in[n]$, and edge $(r,v)$ has length $\sqrt{n}$. There
are $m=n+1$ scenarios: scenario $S_0=\{v\}$ occurs with $1-\frac1n$ probability; and for each $i\in[n]$, scenario
$S_i=\{v,u_i\}$ occurs with $\frac1{n^2}$ probability. The optimal \isoprob value for this instance is $\Omega(n)$ and any reasonable solution clearly will not visit vertex $v$:  it appears in all scenarios and hence provides no information. So if we first follow such an \isoprob solution, the
arrival time for $v$ is $\Omega(n)$; 
since $S_0=\{v\}$ occurs with $1-o(1)$ probability, the
resulting expected latency is $\Omega(n)$. However, the \strp solution that first visits $v$, and then vertices
$\{u_1,\cdots,u_n\}$ has expected latency $O(\sqrt{n})$. 
\end{example}
\smallskip

On the other hand, one can not ignore
the ``isolation aspect'' in \strp either.

\smallskip\begin{example}
Consider another instance of \strp 
on a star-metric with center $r$ and leaves $\{v_i\}_{i=1}^n\cup \{u_i\}_{i=1}^n$. For each $i\in[n]$, edge $(r,v_i)$ has unit length and edge $(r,u_i)$ has length $n$. There
are $n$ scenarios: for each $i\in[n]$, scenario $S_i=\{v_i,u_i\}$ occurs with $\frac1n$ probability. The optimal values for both \strp and \isoprob are $\Theta(n)$. Moreover, any reasonable \strp solution will involve first isolating the realized scenario (by visiting vertices $v_i$s).
\end{example}

\smallskip

Hence, the algorithm needs to interleave the two goals of isolating scenarios and visiting high-probability vertices. This will become clear in the construction of the  latency group Steiner  instances used by our algorithm (Step~\ref{step:trpalg1} in Algorithm~\ref{alg:trppalg}).

\paragraph{Algorithm Outline} Although we can not reduce \strp to \isoprob, we are still able to use ideas from the \isoprob algorithm. The \strp algorithm also follows an iterative approach and maintains a candidate set  $M\sse [m]$ containing the realized scenario. 
We also associate conditional probabilities $q_i:=\frac{p_i}{\sum_{j\in M} p_j}$ for each scenario $i\in M$. In each iteration, the algorithm eliminates a constant fraction of scenarios from $M$: so the number of iterations will be $O(\log m)$. Each iteration involves solving  an instance of the 
{\em latency group Steiner} (\lgs) problem: recall Definition~\ref{def:lgst} and the $O(\log^2n)$-approximation algorithm for \lgs (Corollary~\ref{cor:lgs}). The construction of this \lgs instance is the main point of difference from the \isoprob algorithm. Moreover, we will show that the {\em expected latency} incurred in each iteration is $O(\log^2n)\cdot \opt$. Adding up the latency over all iterations, would yield an $O(\log^2n\,\log m)$-approximation algorithm for \strp.

\paragraph{Using \lgs to partition scenarios $M$} In each iteration, the algorithm formulates an \lgs instance and computes an $r$-tour $\tau$ using Corollary~\ref{cor:lgs}. The details are in Algorithm~\ref{alg:trppalg} below. An important property of this tour $\tau$ is that the number of  candidate scenarios after observing demands on $\tau$ will be at most $|M|/2$ (see Claim~\ref{cl:trp-partn}).

Given a candidate set $M$ of scenarios, it will be convenient to partition the vertices into two parts: $H$ consists of vertices which occur in more than half the scenarios, and $L:=V\setminus H$ consists of vertices occurring in at most half the scenarios. In the \lgs instance (Step~\ref{step:trpalg1} below), we 
introduce $|S_i\cap H|+1$ groups (with suitable weights) corresponding to each scenario $i\in M$.

\begin{algorithm}[h!]
  \caption{$\palgl(\;\langle M,\{q_i\}_{i\in M}, \{S_i\}_{i\in M} \rangle\;)$}
  \label{alg:trppalg}
  \begin{algorithmic}[1]
    \STATE\label{step:trppalg0} {\bf define} $F_v:=\{i\in
    M\mid v\in S_i\}$  for each $v\in V$.

    \STATE {\bf let} 
     $L := \left\{u \in V : |F_u| \leq \frac{|M|}{2} \right\}$, $H := V \setminus
    L$,    and {\small $D_v := \left\{\begin{array}{ll} F_v & \mbox{ if }v\in L\\
    M\setminus F_v &\mbox{ if }v\in H\end{array}\right.$}

   \STATE\label{step:trpalg1} {\bf define} instance ${\cal G}$ of \lgs (Definition~\ref{def:lgst}) 
    on metric $(V,d)$, root $r$ and the following groups:\\
   for each scenario $i\in M$,
   \begin{itemize}
   \item[-] the \emph{main} group $X_i$ 
     of scenario $i$  has weight $|S_i \cap L| p_i$
     and vertices $(L \cap S_i) \cup (H \setminus S_i)$.

   \item[-] for each $v \in S_i \cap H$, group $Y_i^v$
     has weight $p_i$ and vertices $\{v\} \cup  (L \cap S_i) \cup (H \setminus S_i)$.
   \end{itemize}

   \STATE\label{step:trppalg3} \textbf{run} the \lgs algorithm (from Corollary~\ref{cor:lgs}) on instance ${\cal G}$.   \\
   ~~~~~\textbf{let} $\tau :=\langle r,v_1,v_2,\cdots,v_{t-1},r\rangle$ be the $r$-tour  returned.

    \STATE \label{step:trppalg4} \textbf{let} $\{P_k\}_{k=1}^t$ be the
    partition of $M$ where
    {\small $P_k:= \left\{
      \begin{array}{ll}
        D_{v_k}\setminus \left(\cup_{j<k} \, D_{v_j}\right) & \text{if }
        1\le k\le t-1\\
        M\setminus \left(\cup_{j<t} \, D_{v_j}\right) & \text{if }
        k=t\end{array}\right.
    $}

   \STATE \textbf{return} tour $\tau = \langle r,v_1,v_2,\cdots,v_{t-1},r\rangle$ and   partition
   $\{P_k\}_{k=1}^{t}$.
 \end{algorithmic}
\end{algorithm}

\begin{claim}\label{cl:trp-partn}
When $|M|\ge 2$, partition $\{P_k\}_{k=1}^t$ returned by \palgl satisfies $|P_k|\le |M|/2,\, \forall k\in[t]$.
\end{claim}
\begin{pf}
For each $k\in[t-1]$, we have $P_k\sse D_{v_{k}}$ and so $|P_k|\le |M|/2$. We now show that $|P_t|\le 1$ which would
prove the claim. Let $V(\tau)=\{v_1,\ldots,v_{t-1}\}$ denote the vertices visited in the tour $\tau$ output by \palgl.
Consider any $i\in P_t$: we will show that it is unique. By definition of $P_t$, we have $i\not\in \bigcup_{k=1}^{t-1}
D_{v_k}$. By the definition of group $X_i$ and sets $D_v$s, this means that $X_i$ is {\em not covered} by $V(\tau)$.
Since $\tau$ is a feasible solution to ${\cal G}$, $X_i$'s weight must be zero, i.e. $|S_i\cap L|=0$. Thus we have
$S_i\sse H$. Furthermore, if $v_k\in H\setminus S_i$ for any $k\in[t-1]$ then $i\in D_{v_k}$, which implies
$i\not\in P_t$; so $H\cap V(\tau) \, \sse S_i$. Note that each $Y^v_i=\{v\}\cup X_i$ (for $v\in H\cap S_i=S_i$)
must be covered by $\tau$, since $Y^v_i$s have weight $p_i>0$. Also since $X_i$ is not covered by $V(\tau)$, we must
have $v\in V(\tau)$ for all $v\in S_i$. Thus we have $S_i\sse H\cap V(\tau)$, and combined with the earlier
observation, $H\cap V(\tau)\, = S_i$. This determines $i\in M$ uniquely, and so $|P_t|=1\le |M|/2$.
\end{pf}

\paragraph{Final \strp algorithm and analysis}
Given the above partitioning scheme, Algorithm~\ref{alg:trpalg} describes the overall \strp algorithm in a recursive manner.

\begin{algorithm}[!h]
  \caption{$\strp\langle M,\{q_i\}_{i\in M},\{S_i\}_{i\in M} \rangle$}
  \label{alg:trpalg}
  \begin{algorithmic}[1]
    \STATE If $|M|=1$, visit the vertices in this scenario using the $O(1)$-approximation algorithm~\cite{fhr}
    for deterministic traveling repairman, and quit.

    \STATE \label{step:trp2} \textbf{run} $\palgl\langle M,\{q_i\}_{i\in
      M} \rangle$ \\
    ~~~~~\textbf{let} $\tau=(r,v_1,v_2,\cdots,v_{t-1},r)$ be the
    $r$-tour and $\{P_k\}_{k=1}^t$ be the partition of~$M$ returned.

    \STATE \textbf{let} $q'_j:=\sum_{i\in P_k} q_i$ \textbf{for all} $j
    = 1\ldots t$.

    \STATE \label{step:trp3} \textbf{traverse} tour $\tau$ and return
    directly to $r$ after visiting the first vertex $v_{k^*}$
    (for $k^* \in [t]$) that determines that the realized scenario is
    in $P_{k^*} \sse M$.
    
    \STATE {\bf update} the scenarios in $P_{k^*}$ by removing vertices visited in $\tau$ until $v_{k^*}$, i.e. $$S'_i\gets S_i\setminus \{v_1,\ldots,v_{k^*}\},\quad  \mbox{ for all }i\in P_{k^*}.$$

    \STATE  \textbf{run} $\strp\langle P_{k^*},
    \{\frac{q_i}{q'_{k^*}}\}_{i\in P_{k^*}},\{S'_i\}_{i\in P_{k^*}}\rangle$ to recursively cover the
    realized scenario within $P_{k^*}$.
  \end{algorithmic}
\end{algorithm}

The analysis for this algorithm is similar to that for the isolation problem (Section \ref{subsec:iso-alg})
and we follow the same outline.  For any sub-instance $\js$ of \strp, let $\opt(\js)$ denote its optimal value.
Just as in the isolation case (Claim~\ref{cl:alg3}), it can be easily seen that the latency objective function is also
sub-additive.
\begin{claim}\label{cl:trp-subadd}
  For any sub-instance $\langle M, \{q_i\}_{i\in M}, \{S_i\}_{i\in M}\rangle$ and any partition  $\{P_k\}_{k=1}^t$ of $M$,
  \begin{gather}
    \ts \sum_{k=1}^t q'_k \cdot \opt(\langle P_k,
    \{\frac{q_i}{q'_k}\}_{i\in P_k}, \{S_i\}_{i\in P_k}\rangle)\quad \le \quad \opt(\langle M,  \{q_i\}_{i\in M}, \{S_i\}_{i\in M}\rangle),
 \end{gather}
  where $q'_k = \sum_{i\in P_k} q_i$ for all $1\le k\le t$.
\end{claim}

The next property we show is that the optimal cost of the \lgs instance ${\cal G}$ considered in
Steps~\eqref{step:trpalg1}-\eqref{step:trppalg3} of Algorithm~\ref{alg:trppalg} is not too high.

\begin{lemma} \label{lem:latbound}
  For any instance $\js = \langle M, \{q_i\}_{i\in M}, \{S_i\}_{i\in M}\rangle$ of \strp, the
  optimal value of the latency group Steiner instance ${\cal G}$  in
  Step~\ref{step:trppalg3} of Algorithm $\palgl(\js)$ is at most
  $\opt(\js)$.
\end{lemma}

\begin{pf}
Let $T$ be an optimal decision tree for the given  \strp instance $\js$. Note that any internal node of $T$, labeled
$v$, has two children corresponding to the realized scenario being in $F_v$ ({\em yes} child) or $M\setminus F_v$ ({\em
no} child). Now consider the root-leaf path in $T$ (and corresponding tour $\sigma$ in the metric) which starts at $r$,
and at any internal node $v$, moves on to the \emph{no} child if $v \in L$, and moves to  the \emph{yes} child if $v
\in H$. We claim that this tour is a feasible solution to ${\cal G}$, the latency group Steiner instance ${\cal G}$.

To see why, first consider any scenario $i\in M$ that branched off from path  $\sigma$ in decision-tree $T$; let $v$ be
the vertex where the tree path of scenario $i$ branched off from $\sigma$.  If $v \in L$ then by the way we defined
$\sigma$, it follows the ``no'' child of $v$, and so  $v\in S_i \cap L$.  On the other hand, if $v \in H$, then it
must be that $v \in H \setminus S_i$ (again from the way $\sigma$ was defined). In  either  case, $v\in (S_i \cap
L)\cup (H \setminus S_i)$, and hence visiting $v$ covers {\em all} groups, associated with scenario $i$, i.e. $X_i$ and $\{Y_i^v\mid v\in S_i\cap H\}$. Thus $\sigma$ covers all groups of all the scenarios that branched off it in $T$.

Note that there is exactly one scenario (say $a\in M$) that does not branch off $\sigma$; scenario $a$ traverses
$\sigma$ in $T$.
  Since $T$ is a feasible
  solution for \strp, $\sigma$ must visit every vertex in $S_a$.
  Therefore $\sigma$ covers all the groups associated with
  scenario $a$: clearly $\{Y_a^v\mid v\in S_a\cap H\}$ are covered; $X_a$ is also
  covered unless $S_a\cap L=\emptyset$ (however in that case group $X_a$ has zero weight and does not need to be covered- see Definition~\ref{def:lgst}). Thus $\sigma$ is a
  feasible solution to ${\cal G}$.

We now bound the latency cost of tour $\sigma$ for instance ${\cal G}$. In path $\sigma$, let $\alpha_i$ (for each
$i\in M$) denote the coverage time for group $X_i$,
  and $\beta_i^v$ (for $i\in M$ and $v\in S_i\cap H$) the coverage time for group $Y_i^v$.
  The next claim shows that the latency of $\sigma$ for instance ${\cal G}$ is at most $\opt(\js)$.

\begin{claim} The expected cost of $T$, $\opt(\js)\ge \sum_{i\in M} p_i\cdot |L\cap S_i|\cdot \alpha_i + \sum_{i\in
M} \sum_{v\in S_i\cap H} p_i\cdot \beta_i^v$, which is exactly the latency of tour $\sigma$ for the latency group
Steiner instance ${\cal G}$.
\end{claim}
\begin{pf}
Fix any $i \in M$; let $\sigma_i$ denote the shortest prefix of $\sigma$ containing a vertex from $X_i$. Note that by
definition, $\sigma_i$ has length $\alpha_i$. We will lower bound separately the contributions of $S_i\cap L$ and
$S_i\cap H$ to the cost of $T$.

As all but the last vertex in $\sigma_i$ are from $(L\setminus S_i)\cup (H\cap S_i)$, by definition of $\sigma$,
the path $T_{S_i}$ traced in the decision-tree $T$ when scenario $i$ is realized, agrees with this prefix $\sigma_i$. Moreover,
no vertex of $S_i\cap L$ is visited before the end of $\sigma_i$. So under scenario $S_i$, the total arrival time for vertices $L\cap S_i$ is at least $|L\cap S_i|\cdot \alpha_i$. Hence $S_i\cap
L$ contributes  at least $p_i\cdot|L\cap S_i|\cdot \alpha_i$ towards $\opt(\js)$.

Now consider some vertex $v\in S_i\cap H$; let $\sigma_i^v$ denote the shortest prefix of $\sigma$ containing a
$Y_i^v$-vertex. Note that $\sigma_i^v$ has length $\beta_i^v$, and it is a prefix of $\sigma_i$ since $Y^v_i\supseteq
X_i$. As observed earlier, the path traced in decision tree $T$ under scenario $i$  contains $\sigma_i$: so vertex
$v$ is visited (under scenario $i$) only after tracing path $\sigma_i^v$. So the contribution of $v$ (under scenario
$i$) to $\opt(\js)$ is at least $p_i\cdot\beta_i^v$, i.e. the contribution of $S_i\cap H$ is at least $\sum_{v\in
S_i\cap H} p_i\cdot \beta_i^v$
\end{pf}

Thus we have demonstrated a feasible solution to ${\cal G}$ of  latency at most $\opt(\js)$.
\end{pf}

It remains to bound the expected additional latency incurred in Step~\ref{step:trp3} of Algorithm~\ref{alg:trpalg} when
a random scenario is realized. Below we assume a $\rho=O(\log^2n)$ approximation algorithm for latency group Steiner tree (from Corollary~\ref{cor:lgs}).

\begin{lemma}\label{lem:alg3}
  At the end of Step~\ref{step:trp3} of $\strp\langle M, \{q_i\}_{i\in
    M}, \{S_i\}_{i\in  M}\rangle$, the realized scenario lies in $P_{k^*}$. 
    The expected increase in latency due to  this step is at most $2\,\rho \cdot \opt(\langle M, \{q_i\}_{i\in M}, \{S_i\}_{i\in M}\rangle)$.
\end{lemma}
\begin{pf}
The proof that the realized scenario always lies in the $P_{k^*}$ determined in Step~\ref{step:trp3} is identical to that
in Claim~\ref{cl:alg2} of the \isoprob algorithm, and is omitted. We now bound the expected latency incurred. In the solution  $\tau$
to the latency group Steiner instance ${\cal G}$, define $\alpha_i$ as the coverage time for group $X_i$, $\forall i\in
M$; and $\beta_i^v$ as the coverage time for group $Y_i^v$, $\forall i\in M$ and $v\in S_i\cap H$.

Let $i$ denote the realized scenario. Suppose that $k^*=\ell\le t-1$ in Step~\ref{step:trp3}. Then by definition of the
parts $P_k$s, we have $v_{\ell}\in X_{i} = (S_i  \cap L) \cup (H \setminus S_i)$ and $X_i\bigcap
\{v_1,\ldots,v_{\ell-1}\}=\emptyset$. So the length along $\tau$ until $v_\ell$ equals $\alpha_i$. Moreover the total
length spent in this step is at most $2\cdot \alpha_i$, to travel till $v_\ell$ and then return to $r$ (this uses the symmetry and triangle-inequality properties of the metric). So the latency
of any $S_i$-vertex increases by at most this amount. Furthermore we claim that the latency of any $v\in S_i\cap H$ increases by
at most $2\cdot \beta_i^v$: this is clearly true if $\beta_i^v=\alpha_i$; on the other hand if $\beta_i^v<\alpha_i$
then $v$ is visited before $v_\ell$ and so it only incurs latency $\beta_i^v$. So the increase in latency of $S_i$ is
at most $2\sum_{v\in S_i\cap H} \beta_i^v + 2\cdot |S_i\cap L|\,\alpha_i$.

If $k^*=t$ then by the proof of Claim~\ref{cl:trp-partn} the realized scenario $i$ satisfies: $S_i\sse H$, group $X_i$ is not visited by $\tau$ (so $\alpha_i$ is undefined), and all of $S_i$ is visited by $\tau$. In this case the total
latency of $S_i$ is $\sum_{v\in S_i\cap H} \beta_i^v$ which is clearly at most $2\sum_{v\in S_i\cap H} \beta_i^v +
2\cdot |S_i\cap L|\,\alpha_i$; note that $|S_i\cap L|=0$ here.

Thus the expected latency incurred in Step~\ref{step:trp3} is at most $2 \sum_{i\in  M} p_i \cdot \left[ |S_i\cap
L|\,\alpha_i + \sum_{v\in S_i\cap H} \beta_i^v \right]$ which is twice the latency of $\tau$ for the latency group
Steiner instance ${\cal G}$. Finally, since $\tau$ is a $\rho$-approximate solution to ${\cal G}$ and using
Lemma~\ref{lem:latbound}, we obtain the claim.
\end{pf}

Finally, combining Claim~\ref{cl:trp-partn}, Lemma~\ref{lem:alg3} and Claim~\ref{cl:trp-subadd}, 
by a proof identical to that of Theorem~\ref{thm:isolation}, it follows that the final \strp solution has cost $O(\log^2n\,\log m)\cdot \opt$. This completes the proof of 
Theorem~\ref{thm:main3}. 
\medskip

We note that for the \strp problem on metrics induced by a tree, our algorithm achieves an $O(\log n\,\log m)$ approximation ratio (the guarantees in Theorem~\ref{thm:gso} and Corollary~\ref{cor:lgs} improve by a logarithmic  factor on tree metrics). There is also an $\Omega(\log^{1-\epsilon} n)$-hardness of
approximation the \strp problem on tree metrics~\cite{Vish-thesis}. So there is still a logarithmic gap between the best upper and lower bounds for the \strp problem on tree metrics. In going from tree metrics to general, we lose another logarithmic  factor in the approximation ratio.


\section{Concluding Remarks}

In this paper, we studied the problem of constructing optimal decision
trees;  this widely studied problem was previously known to admit logarithmic
approximation algorithms for the case of uniform costs or uniform
probabilities. The greedy algorithms used in these cases do not extend
to the case of non-uniform costs and probabilities, and
we gave a new algorithm that seeks to be greedy with respect to two
different criteria; our $O(\log m)$-approximation is
asymptotically optimal. We then considered a generalization to the adaptive traveling salesman
problem, and obtained an
$O(\log^2 n \log m)$-approximation algorithm for this adaptive TSP problem. We also showed
that any asymptotic improvement on this result would imply an improved
approximation algorithm for the group Steiner tree problem, which is a long-standing
open problem. Finally, we gave an $O(\log^2 n \log m)$-approximation algorithm for the adaptive
traveling repairman problem--- closing the 
 gap
between the known upper and lower bounds in this case remains an 
interesting open problem.

\section*{Acknowledgments.}
A preliminary version appeared in the proceedings of the International Colloquium on Automata, Languages and Programming (ICALP), 2010. We thank Ravishankar Krishnaswamy for many useful conversations; the results on the adaptive traveling repairman problem were obtained in joint discussions, and we thank him for permission to include the results here. We also thank the MOR referees for helpful suggestions that improved the presentation of the paper. 
A. Gupta's research was supported in part by NSF awards CCF-0448095 and CCF-0729022, and an Alfred P.~Sloan Fellowship. R. Ravi's  research was supported in part by NSF grant
  CCF-0728841.

\bibliographystyle{alpha} 
\bibliography{stsp} 

\appendix

\section{Hardness of Approximation for \stsp}
\label{app:hardness}

We show that \stsp is at least as hard to approximate as group Steiner tree.
\begin{theorem}\label{th:iso-hard}
If there is an $\alpha$-approximation algorithm for \stsp then there is an  $(\alpha+o(1))$-approximation algorithm for group Steiner tree.
Hence \stsp is $\Omega(\log^{2-\epsilon}n)$ hard to approximate even on tree metrics.
\end{theorem}
\begin{pf}
This reduction is similar to the reduction~\cite{CPRAM11} from set-cover to the \dtp; we give a proof in
context of \stsp for completeness.

  Consider an arbitrary instance of group Steiner tree on metric $(V,d)$
  with root $r$ and groups $X_1,\cdots,X_g\sse V$; let $\opt$ denote its
  optimal value. Assume without loss of generality that $X_i\ne X_j$ for
  all $i\ne j$, and the minimum non-zero distance in $d$ is one. We construct an instance of \stsp as follows. Let
  $V'=V\cup\{s\}$ where $s$ is a new vertex (representing a copy of $r$), and define metric $d'$ on
  $V'$ as:
  $$d'(u,v) :=\left\{
    \begin{array}{ll}
      d(u,v) & \mbox{ for }u,v \in V\\
      d(u,r) & \mbox{ for }u\in V,~v=s
    \end{array}\right.,\qquad \forall (u,v)\in {V'\choose 2}
  $$

There are $g+1$ scenarios in the \stsp instance: $S_i:=X_i\cup \{s\}$ for $i\in[g]$,
  and $S_{g+1}:=\{s\}$, with probabilities
  $$p_i :=\left\{
    \begin{array}{ll}
      \frac1{gL} & \mbox{ if }1\le i\le g\\
      1-\frac1L & \mbox{ if }i=g+1
    \end{array}\right.,
  $$
Above $L\gg 2 n\cdot \max_{u,v} d(u,v)$ is some large value. The root in the \stsp instance remains $r$. Let $\opt'$
denote the optimal value time of this instance. We will show that $(1-o(1))\cdot \opt\le \opt'\le \opt+1$ which would
prove the  theorem.

\noindent {\bf (A) $\left( 1-\frac1L\right) \opt \, \le \, \opt'$.} Consider the optimal solution to the \stsp
instance; let $\sigma$ denote the $r$-tour traversed by this decision tree under scenario $S_{g+1}$. We now argue that
$\sigma$ is a feasible solution to the group Steiner  tree instance, i.e., $\opt\le d(\sigma)$. Suppose for a
contradiction that $\sigma$ does not visit any $X_i$-vertex for some $i\in[g]$. Then observe that the $r$-tour
traversed by this decision tree under scenario $S_i$ is also $\sigma$, since the  decision tree can not distinguish scenarios
$S_i$ and $S_{g+1}$ (the only way to do this is by visiting some $X_i$-vertex). However this violates the requirement
that the tour (namely $\sigma$) under scenario $S_i$ must visit all vertices $S_i\supseteq X_i$. Finally, we have
$\opt'\ge (1-\frac1L)\cdot d(\sigma)\ge \left( 1-\frac1L\right) \opt$ as required. 

\noindent {\bf (B) $\opt'\le \opt+1$.} Let $\tau$ denote an optimal
  $r$-tour for the given \mgs instance, so $d(\tau)=\opt$. Consider the
  following solution for \stsp:
  \begin{enumerate}
  \item Traverse $r$-tour $\tau$ to determine whether or not $X_{g+1}$ is
    the realized scenario.
  \item If no demands observed on $\tau$ (i.e. scenario $S_{g+1}$ is realized), visit vertex $s$ and stop.
  \item If some demand observed on $\tau$ (i.e. one of scenarios $\{S_i\}_{i=1}^g$ is realized), then visit {\em all} vertices in $V$ along an
    arbitrary $r$-tour and stop.
   \end{enumerate}
It is clear that this  decision tree is feasible for the \stsp instance.  For any $i\in [g+1]$, let $\pi_i$ denote the
$r$-tour traversed under scenario $S_i$ in the above \stsp  decision tree. We have
  $d(\pi_{g+1})=d(\tau)\le \opt$, and
  $d(\pi_{i})\le 2n\cdot \max_{u,v} d(u,v) \le L$ for all $i\in [g]$. Thus the resulting \stsp objective is
  at most:
  $$\left(1-\frac1L\right)\cdot \opt + g\cdot \frac1{gL}\cdot L \le
  \opt + 1$$
  Thus we have the desired reduction.
\end{pf}

\end{document}